\documentclass[12pt]{article}
\usepackage{amssymb,amsmath,amscd}
\usepackage{epsfig}
\usepackage{epstopdf}
\usepackage{latexsym}
\usepackage{graphicx}
\usepackage{booktabs}
\usepackage{bbm}
\usepackage[english]{babel}
\usepackage{blindtext}
\usepackage{arydshln}
\usepackage{multirow}
\usepackage{float}
\usepackage{tikz}
\usepackage{tikz-cd}
\usetikzlibrary{arrows.meta}
\usetikzlibrary{matrix,arrows,backgrounds,fit,chains,intersections}
\usetikzlibrary{decorations.pathreplacing,decorations.markings}
\usetikzlibrary{calc}
\usepackage{youngtab}
\usepackage{titlesec}
\usepackage[T1]{fontenc}
\usetikzlibrary{positioning}
\usepackage{xcolor}
\usepackage{datetime}
\usepackage{skak}
\usepackage{staves}

\usepackage{cite}
\usepackage{hyperref}
\usepackage{setspace}
\hypersetup{
	colorlinks=true,
	linkcolor=dark-blue,
	citecolor=dark-red,
	urlcolor=dark-green,
	linktoc=page
}

\usepackage[hmarginratio=1:1,top=32mm,columnsep=20pt]{geometry} 
\usepackage{fancyhdr} 
\numberwithin{equation}{section}

\titleformat{\section}[block]{\large\bfseries\centering}{\thesection}{1em}{} 
\titleformat{\subsection}[block]{\bfseries}{\thesubsection}{1em}{} 

\tikzset{->-/.style={decoration={
			markings,
			mark=at position .5 with {\arrow{stealth'}}},postaction={decorate}}}
\tikzset{-<-/.style={decoration={
					markings,
					mark=at position .5 with {\arrow{stealth' reversed}}},postaction={decorate}}}
\tikzset{
	position label/.style={
		below = 3pt,
		text height = 2ex,
		text depth = 1ex
	},
	brace/.style={
		decoration={brace, mirror},
		decorate
	}
}
\tikzset{
	state/.style={
		rectangle,
		rounded corners,
		draw=black, very thick,
		minimum height=2em,
		inner sep=2pt,
		text centered,
	},
}

\definecolor{dark-gray}{gray}{0.20}
\definecolor{gray}{gray}{0.30}
\definecolor{light-gray}{gray}{0.80}
\definecolor{dark-red}{rgb}{0.7,0,0}
\definecolor{dark-green}{rgb}{0.1,0.4,0}
\definecolor{dark-blue}{rgb}{0.3,0.3,0.7}
\definecolor{light-blue}{rgb}{0.8,0.8,1}
\definecolor{cardinal}{rgb}{0.6,0,0}
\definecolor{darkgreen}{rgb}{0,0.5,0}
\definecolor{golden}{rgb}{0.92, 0.7, 0}
\definecolor{midnight}{rgb}{0, 0, 0.5}
\definecolor{darkblue}{rgb}{0.2, 0, 0.8}

\def\rmi{{\rm i}}

\def\Tr{{\rm Tr}\,}
\def\tr{{\tt Tr}\,}
\newcommand{\dvol}{\mathrm{vol}}
\newcommand{\vol}{\mathrm{V}}
\newcommand{\dd}{\mathrm{d}}
\newcommand{\e}{\mathrm{e}}
\newcommand{\f}[2]{\frac{#1}{#2}}
\newcommand{\sgn}{\mathrm{sgn}}
\newcommand{\lcm}{\mathrm{lcm}}

\def\SL{{\rm SL}}

\def\SO{{\rm SO}}
\def\U{{\rm U}}
\def\SU{{\rm SU}}
\def\Sp{{\rm Sp}}
\def\PSL{{\rm PSL}}

\def\su{\mathfrak{su}}

\title{\vspace{-10mm}\fontsize{22pt}{10pt}\selectfont\textbf{Wrapped Branes and Punctured Horizons}\vspace{10mm}}

\author{Nikolay Bobev$^\text{\runictext{a}}$, Pieter Bomans$^\text{\runictext{a}}$ and Fri{\dh}rik Freyr Gautason$^\text{\runictext{a},\runictext{b}}$\\[5mm] 
	\normalsize $^\text{\runictext{a}}$Instituut voor Theoretische Fysica, KU Leuven\\
\normalsize Celestijnenlaan 200D, 3001 Leuven, Belgium\\[3mm]
\normalsize $^\text{\runictext{b}}$University of Iceland, Science Institute\\
\normalsize Dunhaga 3, 107 Reykjav{\'i}k, Iceland\\[3mm]
	\texttt{\small\href{mailto:nikolay.bobev@kuleuven.be}{\{nikolay.bobev}, \href{mailto:pieter.bomans@kuleuven.be}{pieter.bomans}, \href{mailto:ffg@kuleuven.be}{ffg\}@kuleuven.be}}
}
\date{}

\begin{document}  
 
\maketitle
\thispagestyle{fancy}

\vspace{10mm}

\begin{abstract}

\noindent 

Large classes of AdS$_p$ supergravity backgrounds describing the IR dynamics of $p$-branes wrapped on a Riemann surface are determined by a solution to the Liouville equation. The regular solutions of this equation lead to the well-known wrapped brane supergravity solutions associated with the constant curvature metric on a compact Riemann surface. We show that some singular solutions of the Liouville equation have a physical interpretation as explicit point-like brane sources on the Riemann surface. We uncover the details of this picture by focusing on $\mathcal{N}=1$ theories of class $\mathcal{S}$ arising from M5-branes on a punctured Riemann surface. We present explicit AdS$_5$ solutions dual to these SCFTs and check the holographic duality by showing the non-trivial agreement of 't Hooft anomalies.

\end{abstract}

\noindent 

\vfill

\thispagestyle{empty}

\newpage

\setcounter{tocdepth}{2}
\tableofcontents

\newpage

\section{Introduction} \label{sec:Intro}

Studying the low-energy physics of $p$-branes in string and M-theory wrapped on an $n$-dimensional curved manifold, $\mathcal{M}$, has provided a rich arena for understanding the dynamics of strongly coupled quantum field theories. This is facilitated by three distinct vantage points that provide complementing insights into the physics of these systems. One can view this setup as realizing a partial topological twist on $\mathcal{M}$ of the $(p+1)$-dimensional supersymmetric QFT on the  world-volume of the brane. At low energies this leads to a QFT in $(p-n+1)$ dimensions which preserves part of the original supersymmetry. Alternatively, one can realize the same system more geometrically by studying the low-energy dynamics of the $p$-brane wrapped on a calibrated cycle $\mathcal{M}$ in a special holonomy manifold. Holography offers a third point of view on the same physics. When the number of $p$-branes is large they backreact on the geometry and this often leads to supergravity solutions dual to the QFTs of interest. In this work we will study the case $n=2$, i.e. $\mathcal{M}$ is a Riemann surface, and show how to construct these supergravity solutions for various $p$-branes in the presence of punctures on the Riemann surface.

Studying wrapped branes on Riemann surfaces using holography was initiated in the seminal work of Maldacena-N\'u\~nez \cite{Maldacena:2000mw}, see \cite{Gauntlett:2003di} for a review. Renewed interest in the physics of these system arose from understanding the four-dimensional $\mathcal{N}=2$ quantum field theories of class $\mathcal{S}$ arising on worldvolume of M5-branes wrapping a Riemann surface \cite{Gaiotto:2009we,Gaiotto:2009hg}. A key role in the class $\mathcal{S}$ construction is played by the punctures on the Riemann surface, $\mathcal{C}$. The encode information about additional flavor symmetries and matter fields in the quantum field theory. In the brane setup these punctures correspond to M5-branes which intersect the Riemann surface at points and share four dimensions with the wrapped branes. Such punctures on the Riemann surface can also be incorporated in the holographic description of the class $\mathcal{S}$ setup. It was shown in \cite{Gaiotto:2009gz} that the gravitational description this system is captured by a generalization of the class of $\tfrac{1}{2}$-BPS AdS$_5$ solutions described in \cite{Lin:2004nb}. These solutions are described by a single function obeying the non-linear $\SU(\infty)$ Toda equation which in the presence of punctures on $\mathcal{C}$ is modified by including a singular source. A large number of non-trivial consistency checks of this proposal were performed in \cite{Gaiotto:2009gz} and all of them lead to nice agreement with the field theory analysis of \cite{Gaiotto:2009we,Gaiotto:2009hg}.

Given this success it is natural to ask whether this supergravity description of wrapped branes on punctured Riemann surfaces can be generalized to branes in other dimensions or systems with smaller number of supercharges. Unfortunately the approach followed in \cite{Gaiotto:2009gz} proves to be hard to generalize in these setups. For instance if one wants to generalize the $\mathcal{N}=1$ class $\mathcal{S}$ construction of \cite{Benini:2009mz,Bah:2011vv,Bah:2012dg} to Riemann surfaces with punctures one has to study $\tfrac{1}{4}$-BPS AdS$_5$ backgrounds of eleven-dimensional supergravity. While solutions of this type have been classified in \cite{Gauntlett:2004zh} the supergravity BPS equations reduce to a complicated system of coupled nonlinear PDEs. Despite the progress described in \cite{Bah:2013qya,Bah:2015fwa}, it appears to be hard to apply the idea of \cite{Gaiotto:2009gz} and introduce singular brane sources to this system of equations and find explicit solutions. This state of affairs is even more grim for other wrapped brane systems like D3- or M2-branes. Where one should classify $\tfrac{1}{8}$-BPS AdS$_3$ or $\tfrac{1}{16}$-BPS AdS$_2$ solutions of type IIB or eleven-dimensional supergravity and introduce singular sources to the corresponding BPS equations. Given this impasse it is clearly beneficial to explore alternative approaches to the construction of this type of supergravity solutions. Our goal here is to present one such approach.

Our strategy is based on the observation that the constructions of many wrapped branes supergravity solutions proceeds along the lines of \cite{Maldacena:2000mw}. Namely, one starts with an appropriate lower-dimensional gauged supergravity which is a consistent truncation of ten- or eleven-dimensional supergravity on a compact manifold (typically a sphere). One then studies an appropriate Ansatz for the fields of the gauged supergravity theory which implements the holographic description of the partial twist of the dual SCFT on the compact Riemann surface and then constructs solutions of the supergravity BPS equations. We modify this procedure in a simple way, we allow for the metric on the Riemann surface to be general, as opposed to the constant curvature metric in \cite{Maldacena:2000mw}. We then study this setup in the maximal gauged supergravity theories in four, five, six and seven dimension relevant for the holographic description of M2-, D3-, D4-D8, and M5-branes. We  find that in all cases the supergravity BPS equations reduce to the following PDE on the Riemann surface
\begin{equation}\label{Liouville}
	\square \varphi + \kappa \e^{\varphi} = 0\,,\notag
\end{equation}
for one of the functions in the Ansatz, where $\kappa$ is the normalized curvature of $\mathcal{C}$. This is the well-known Liouville equation. Its regular solutions lead to the well-known supergravity solutions describing branes wrapped on a compact Riemann surface. Our main observation is that introducing a singular source on the right hand side of the Liouville equation allows for new solutions that have not been explored before. We interpret these solutions as providing the supergravity description of branes wrapped on a punctured Riemann surface. We can then rely on a well-known supergravity uplift formulae to present the corresponding ten- or eleven-dimensional supergravity solutions. In this way we circumvent the need to classify AdS supergravity solutions and solve complicated PDEs directly in the ten or eleven-dimensional theory.

To gain confidence in the proposal described above we study it in detail for the case of M5-branes wrapped on a punctured Riemann surface with a general choice of topological twist preserving $\mathcal{N}=1$ supersymmetry. We first show that for the special twist with $\mathcal{N}=2$ supersymmetry our results are compatible with the ones in \cite{Gaiotto:2009gz}. This comparison also exhibits a small limitation in our approach. The singular source in the Liouville equation does not capture the full information about the puncture present in the eleven-dimensional solutions of \cite{Gaiotto:2009gz} and only serves as an approximate description. Nevertheless, the gauged supergravity approach provides enough information for many interesting questions, in particular in the large $N$ approximation. We show this utility by studying more general $\mathcal{N}=1$ setups of class $\mathcal{S}$ where we show how the results computed using the supergravity solutions agree with the M5-brane anomaly polynomial as well as explicit constructions of the dual quantum field theories using vector and hyper multiplets as well as $\mathcal{T}_N$ building blocks.

In the next section we present our main proposal on how to treat punctured Riemann surfaces in gauged supergravity and present some details on solutions corresponding to wrapped M2-, D3-, D4-D8, and M5-branes. From Section \ref{sec:M5branes} onwards we focus on M5-branes wrapped on punctured Riemann surfaces in order to accumulate evidence for our general prescription. We start by describing the different twists of the $\mathcal{N}=(2,0)$ theory and summarize how to integrate the anomaly polynomial of the M5-branes over the punctured Riemann surface to obtain the anomalies of the IR four-dimensional theories. In Section \ref{sec:punctures} we revisit the AdS$_5$ supergravity solutions of Section \ref{sec:puncbranes} and describe our treatment of singularities on the Riemann surface. We also compute the conformal anomalies of these solutions holographically and obtain an exact match with the result from the anomaly polynomial at leading order in $N$. Moreover, we compute the dimension of protected operators arising from M2-branes wrapping the Riemann surface and describe the marginal deformations of our solutions. Finally, in Section \ref{sec:quivers} we construct the quiver gauge theories dual to a subset of our supergravity solutions. We compute the conformal anomalies, the dimensions of the M2-brane operators and the dimension of the conformal manifold and on all fronts find agreement with supergravity and the anomaly polynomial. We finish by briefly discussing various non-trivial Seiberg-like dualities. The four appendices contain technical details on the supergravity constructions we employ, a review of some solutions of the Liouville equation, as well as a brief summary of our SCFT conventions.

\section{Punctured horizons} 
\label{sec:puncbranes}

We consider $(d-3)$-dimensional SCFTs arising from twisted compactifications of SCFTs in $d-1$ dimensions on a punctured Riemann surface $\mathcal{C}$. These can be realized as the theories living on the worldvolume of D- or M-branes where the worldvolume takes the form $\mathbf{R}^{1,d-4}\times \mathcal{C}$. In general when putting a supersymmetric field theory on a curved manifold all supersymmetries are broken. However, by performing a (partial) topological twist we can preserve some supersymmetry \cite{Witten:1988ze,Bershadsky:1995qy,Maldacena:2000mw}. The generator of supersymmetry is a spinor, $\epsilon$, which in the presence of a background metric and R-symmetry gauge field obeys an equation of the schematic form
\begin{equation}
	\left(\partial_\mu + \f14\omega_\mu^{ab}\gamma_{ab} +A_\mu^I\Gamma_I\right)\epsilon = 0\,.
\end{equation}
Here $\omega_\mu^{ab}$ is the spin connection and $A_\mu$ is the background gauge field coupled to the R-symmetry. By identifying the structure group of $\mathcal{C}$ with a subgroup of the R-symmetry this equation can be solved by taking a constant spinor obeying $\partial_\mu\epsilon = 0$. In order to perform a topological twist on a Riemann surface we need at least $\U(1)_R$ superconformal R-symmetry. In this paper we focus on SCFTs with the maximal number of supercharges in three, four, five, and six dimensions which have a larger R-symmetry group and thus the $(d-3)$-dimensional SCFTs in the IR preserve some amount of supersymmetry.\footnote{The analysis below can be extended to SCFTs and supergravity theories with smaller amount of supersymmetry, see for example \cite{Benini:2015bwz,Bobev:2017uzs}.}

Following the seminal work \cite{Maldacena:2000mw} we study these twisted SCFTs using holography. To this end we consider maximally supersymmetric gauged supergravity theories in $d=4,5,6$ and $7$ dimensions and study the most general topological twist on $\mathcal{C}$ in every dimension. The construction follows a similar pattern for every value of $d$ and before we focus on each individual case we describe the general structure.

We work with a truncation of the maximal gauged supergravity which reduces the bosonic fields to the metric and a number of Abelian gauge fields and real scalars. The supergravity solutions dual to the twisted SCFTs described above are of the following form 
\begin{equation}\label{Ansatz}
\begin{aligned}
	\dd s_{d}^2  =& \e^{2f(r,x_1,x_2)}\left( -\dd t^2 + \dd z_1^2 +\cdots +\dd z_{d-3}^2 + \dd r^2\right) + \e^{\hat{\varphi}(r,x_1,x_2)}\left( \dd x_1^2+\dd x_2^2\right)\,,\\
	A^{(i)} =& A_{x_1}^{(i)}(r,x_1,x_2)\dd x_1 + A_{x_2}^{(i)}(r,x_1,x_2)\dd x_2 \,,\\
	\lambda_I =& \lambda_I(r,x_1,x_2) \,,\hspace{95pt} \,.
\end{aligned}
\end{equation}
The range of the indices $i=1,\dots,n_A$ and $I=1,\dots,n_\lambda$ differs on a case by case basis and will be specified for each dimension separately. In these expressions all functions -- $f$, $\hat{\varphi}$, $A_{x_1,x_2}^{(i)}$ and $\lambda_I$ -- only depend on the radial coordinate $r$ and the coordinates $x_1$ and $x_2$ of the Riemann surface. For Riemann surfaces with Gaussian curvature $\kappa=-1$, the coordinates $(x_1,x_2)$ parametrize the hyperbolic plane $\mathbf{H}$ which we quotient by a discrete Fuchsian subgroup $\Gamma \in \PSL(2,\mathbf{R})$ to obtain a genus $\mathbf{g}>1$ Riemann surface. Furthermore we focus on the IR behaviour of these wrapped brane solutions where the geometry becomes $\text{AdS}_{d-2}\times \mathcal{C}$. The metric functions in this IR region are fixed to
\begin{equation}\label{IRfields}
	\begin{aligned}
	f =&  -\log r + \log\f{2}{g}+f_0\,,\\
	\hat{\varphi} =&\,\varphi(x_1,x_2) + 2\log\f{2}{g}+\varphi_0\,, 
	\end{aligned}
\end{equation}
where $f_0$ and $\varphi_0$ are constants and the scalars take constant values. The gauge coupling $g$ is related to the radius of the UV AdS$_d$ solution, $R_{\text{AdS}_d}=\f{2}{g}$. It is worth pointing out that the more general BPS equations which describe the holographic RG flow from the AdS$_d$ UV region to the AdS$_{d-2}$ IR near-horizon region were studied in detail in \cite{Anderson:2011cz,BGP}. The result is that these equations admit solutions for arbitrary metric on the Riemann surface in the UV region, however in the IR the metric flows to the constant curvature metric on $\mathcal{C}$. This behavior is known as holographic uniformization \cite{Anderson:2011cz,BGP}. From now on we concentrate solely on the IR region and investigate the resulting near-horizon geometries. As described in Appendix \ref{app:BPS} one can show that the BPS equations at the IR fixed point reduce to a number of algebraic equations for the scalars together with one universal second order equation for the conformal factor $\varphi$ of the metric on the Riemann surface. The gauge fields in turn are -- up to a choice of twist -- fully determined in terms of this function $\varphi$. The equation determining $\varphi$ is given by
\begin{equation}\label{Liouville}
	\boxed{~~\square \varphi + \kappa \e^{\varphi} = 0~~}
\end{equation}
This is nothing but the Liouville equation for the conformal factor of the metric on the Riemann surface.\footnote{Some properties of the Liouville equation are summarized in Appendix \ref{app:Liouville}.}. When one considers smooth Riemann surfaces one finds the constant curvature metric on the covering space
\begin{equation}\label{concurv}
	\begin{array}{lll}
	\varphi(x_1,x_2) = -2\log x_2 + \log 2 & \text{for}\qquad & \kappa = -1\,,\\
	\varphi(x_1,x_2) = 0 & \text{for} & \kappa = 0\,,\\
	\varphi(x_1,x_2) = -2\log(1+x_1^2+x_2^2) + \log 8\qquad & \text{for} & \kappa = 1\,.\\
	\end{array}
\end{equation}
The crucial observation for out work is that there are more general solutions to the Liouville equation. One can construct many more IR AdS$_{d-2}$ solutions by allowing singular solutions to the Liouville equation where the Riemann surface includes conical defects or punctures.

Our main observation is that regular punctures on the wrapped curve correspond to conical defects of the Riemann surface in the lower dimensional supergravity description. In order to accommodate such singularities one has to add localized sources to the Liouville equation
\begin{equation}\label{Liouvillesource}
	\square \varphi + \kappa \e^{\varphi} = \sum_{i}4\pi(1-\xi_i)\delta^{(2)}(P_i)\,.
\end{equation}
where $i$ runs over all singularities and $0\leq \xi_i \leq 1$ specifies the opening angle of the conical defect at the point $P_i$. The limiting values $\xi=0$ and $\xi=1$ correspond to respectively a true puncture and a regular point. Near a singular point the Liouville field needs to satisfy appropriate boundary conditions. Near a conical singularity with defect angle $\xi$, the boundary conditions are
\begin{equation}\label{liouvillecondef}
	\varphi = -2(1-\xi)\log r \,,\qquad \text{for } \quad r\rightarrow 0\,.
\end{equation}
where $r^2 = x_1^2 + x_2^2$ and the singularity is chosen to lie at the origin. Once a solution to the Liouville equation on a Riemann surface with prescribed singularities is given, the gauge field strength is fully determined by the conformal factor $\varphi$, up to a choice of partial topological twist. In terms of the spin connection $\omega_\mu = \f12 \omega_\mu^{ab}\epsilon_{ab}$, we choose the R-symmetry background gauge field $A_\mu$ such that its field strength is, up to a $d$-dependant prefactor, given by
\begin{equation}
F_= \sum_i F^{(i)}T_i \sim  -\kappa\,\vol_{\mathbf{g},\xi}^{-1}\, T \, \dd\omega\,.
\end{equation}
for $\kappa\neq 0$ and $F\sim -T \dvol_{\mathbf{g},\xi}$ for $\kappa=0$. Here we defined $\dvol_{\mathbf{g},\xi}$ to be the volume form on the singular Riemann surface of genus $\mathbf{g}$ with $n$ conical defects with opening angles $\xi_j$ and $\vol_{\mathbf{g},\xi}$ is its volume
\begin{equation}
\vol_{\mathbf{g},\xi} = \f{2\pi}{\kappa}\left(2-2\mathbf{g} - \sum_{j=1}^{n}(1-\xi_j)\right)\,.
\end{equation}
This implies that in order guarantee that the volume is positive we must have $\kappa = \sgn[2-2\mathbf{g} - \sum_{j=1}^{n}(1-\xi_j)]$. The gauge field is taken along the generator $T = \sum a_i T_i$ where $T_i$ are the generators of the Cartan of the R-symmetry group. The $a_i$ are constants parametrizing the partial topological twist. In order to preserve supersymmetry $a_i$ need to obey the constraint
\begin{equation}\label{twistcond}
	2\pi\sum_i a_i = -\kappa\,\vol_{\mathbf{g},\xi}\,. 
\end{equation}
For smooth Riemann surfaces the condition for the R-symmetry bundle to be well defined, together with the twisting condition, \eqref{twistcond}, imply that the $a_i$ can only take quantized values $a_i\in \mathbf{Z}$. On the other hand, for a singular Riemann surface with conical singularities, with deficit angles $\{\xi_j\}_{j=1,\cdots,n}$, the quantization condition is slightly altered and becomes
\begin{equation}
	\lcm(\{\xi^{-1}_j\})a_i \in \mathbf{Z}\,.
\end{equation}
This condition corresponds to the quantization of the fluxes $F^{(i)}$ and is very similar in spirit to the quantization of electric charge in the presence of a magnetic monopole. Namely, when a monopole of magnetic charge $m$ is present the charge quantization condition takes the form $m n \in \mathbf{Z}$. 

Another useful point of view on this quantization condition is more geometric and is offered by uplifting these backgrounds to string or M-theory. There the quantization arises for imposing the the normal bundle to the wrapped branes is well-defined. For an $\SU(N)$ gauge group the twisted SCFTs theories describe the low energy limit of a D- or M-brane wrapped on a singular curve $\mathcal{C}$ inside a Calabi-Yau $m$-fold, where $m = n_A + 1$, which is a $\bigoplus_{i=1}^{n_A} \mathcal{L}_i$ line bundle over $\mathcal{C}$. 
\begin{equation}
\begin{tikzcd}
\mathbf{C}^{n_A} \arrow[r,hook]  & \bigoplus_{i=1}^{n_A} \mathcal{L}_i \arrow[d]\\
&\mathcal{C}
\end{tikzcd}
\end{equation}
The degrees of the line bundles are $\deg(\mathcal{L}_i)=-\kappa a_i$ for $\kappa\neq 0$ and $\deg(\mathcal{L}_i)=a_i$ for $\kappa=0$. The Calabi-Yau condition reduces to $\bigotimes_{i=1}^{n_A}\mathcal{L}_i = K_\mathcal{C}$, where $K_\mathcal{C}$ is the canonical line bundle of $\mathcal{C}$. This relation is equivalent to the twist condition \eqref{twistcond}. Note that due to the presence of singularities the degrees of the line bundles can take rational values \cite{Reid:1987}.

\begin{figure}[H]
	\centering
	\includegraphics[scale=0.6]{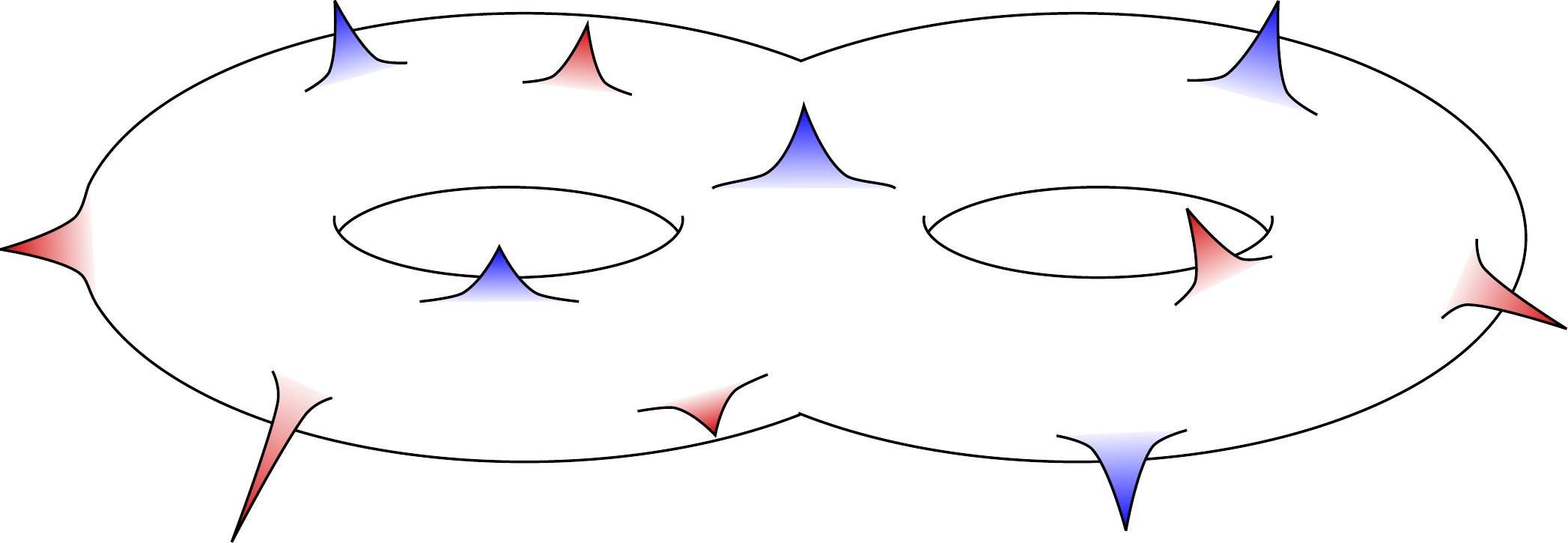}
	\caption{Regular punctures on the wrapped Riemann surface correspond to conical defects on the Riemann surface in the lower dimensional supergravity. The local information of a puncture is encoded in the opening angle of the corresponding conical defect. The color of the shading surrounding the singularity determines in which transverse direction the puncture extends.}
	\label{fig:congen}
\end{figure}

It proves convenient for the subsequent analysis to split the parameters $a_i$ into global and local parts
\begin{equation}\label{eq:aglobloc}
a_i = a^i_{\rm global} + a^i_{\rm local}\,,
\end{equation}
where the global part corresponds to the background flux present for the topological twist on a smooth $\mathcal{C}$, i.e. such that $\sum_i a^i_{\rm global} = (2g-2)$. For each puncture there is a choice to add the local contribution to one of the magnetic fields or equivalently to one of the $a_i$'s indicating in which direction normal to the branes the puncture extends.\footnote{One can consider the more general case where the contribution from a single singularity is split over multiple $a^i$. In this case the singularity still extends in a one-dimensional subspace of the transverse space the will locally still preserve half-maximal supersymmetry. By choosing a different basis for the transverse space we recover the same picture as above. However, when a singularity extends in a subspace of the transverse space with dimension $d>1$ we have a truly different situation and the intuitive picture developed above will no longer be correct. To the best of our knowledge this situation has not been studied in the literature and we do not consider it here.} Graphically we can represent this choice by assigning a color to each conical defect, see Figure~\ref{fig:congen} for an illustration valid for M5-branes wrapped on $\mathcal{C}$, as follows
\begin{equation}
	a^1_{\rm local} = \sum_{\text{red punctures}}(1-\xi_j)\,,\qquad a^2_{\rm local} = \sum_{\text{blue punctures}}(1-\xi_j)\,, \qquad\cdots
\end{equation}
For theories of class $\mathcal{S}$ with gauge group $\SU(N)$ regular punctures are classified by Young tableaux with $N$ boxes \cite{Gaiotto:2009we,Chacaltana:2010ks}. Similarly we conjecture that in other dimensions and with smaller number of supercharges many punctures can be classified in the same manner. Around a puncture we can uplift our supergravity solution to ten or eleven dimensions. For $\xi^{-1}\in \mathbf{Z}$ the uplifted geometry  around the puncture in 10 or 11 dimensions takes the form
\begin{equation}\label{Zksing}
\dd s_{10(11)}^2 \simeq \dd s_{\text{AdS}_{d-1}}^2 + \dd s^2_{\mathbf{R}^4/\mathbf{Z}_{\xi^{-1}}} + \dd s^2_{S^{7(8) - d}}\,.
\end{equation}
In class $\mathcal{S}$ theories this kind of geometry corresponds to a puncture characterized by a rectangular Young tableau with rows of equal length $\xi^{-1}$. The geometrical structure associated to punctures with more general Young tableaux is more complicated as the singularities are spread out along the internal space \cite{Gaiotto:2009gz}. In order to analyse these singularities in full detail one should consider the full ten or eleven-dimensional description of the solution. However, in the large $N$ limit we can approximately describe these solutions with conical singularities with defect angles $\xi\in\{ \frac{r}{N} | r=1,\dots,N-1 \}$. A conical defect with $\xi = \frac{r}{N}$ corresponds to the set of all regular punctures described by Young tableaux with $r$ rows; to select the specific Young tableau from this class one needs additional information encoded in the transverse geometry. Such more general punctures might violate the quantization conditions formulated above. For the moment we do not worry about this and merely consider this as an effective description in the gauged supergravity. Using this approximation we show that for $\mathcal{N}=1$ theories of class $\mathcal{S}$ we can match the conformal anomalies and dimensions of specific operators for all types of punctures in the dual field theory.

In the remainder of this section we construct explicit gauged supergravity solutions corresponding to M2-, D3-, D4-D8- and M5-branes wrapping the singular Riemann surface $\mathcal{C}$. 

\subsection{M2-branes on singular curves}

The gauge theory arising on the worldvolume of $N$ M2-branes is given by the three-dimensional $\mathcal{N}=8$ ABJM theory \cite{Aharony:2008ug} which at large $N$ is dual to eleven-dimensional supergravity on $\text{AdS}_4\times S^7$.\footnote{For simplicity we focus on the ABJM theory with Chern-Simons level $k=1$.} A twisted compactification of this theory on a complex curve $\mathcal{C}$ is described holographically by an eleven-dimensional supergravity background which is asymptotically locally $\text{AdS}_4\times S^7$ but for which the topology at a fixed value of the radial coordinate is an $S_7$ fibration over $\mathcal{C}$. An efficient way to construct these supergravity solutions is to study them in the maximal four-dimensional $\SO(8)$ gauged supergravity \cite{deWit:1982bul} which is a consistent truncation of the eleven-dimensional theory on $S^7$. For our purposes we do not need the full structure of the four-dimensional $\mathcal{N}=8$ theory and restrict to a further truncation studied in \cite{Cvetic:1999xp}. The bosonic subsector of this truncation consists of a metric, four abelian gauge fields in the Cartan of the $\SO(8)$ gauge group and three real neutral scalars. It can been shown that all solutions of this truncation can be uplifted to solutions of eleven-dimensional supergravity \cite{Cvetic:1999xp}.

In \cite{Gauntlett:2001qs,Benini:2015eyy} the near-horizon geometry of M2-branes wrapped on a smooth Riemann surface was analysed using the same truncation. One can show that by inserting our ansatz in the BPS equations they indeed reduce to the Liouville equation for the conformal factor $\varphi$ together with algebraic equations for the other fields \cite{BGP}. In terms of $\varphi$, the gauge field strengths are given by\footnote{Here and in the upcoming cases this expression for the field strengths is valid only when $\kappa=\pm1$. When $\kappa=0$, the field strengths are given by $F^i = \f{a^i}{g}\dd x_1 \wedge \dd x_2$.}
\begin{equation}
	F^i_{(2)} = -\kappa\f{2\pi\,a_i}{g}\vol^{-1}_{\mathbf{g},\xi} \, \dvol_{\mathbf{g},\xi}\,,
\end{equation}
where $g$ is the gauge coupling constant of the supergravity theory and $i=1,2,3,4$. The constants $a_i$ determine the specific choice of twist and are constraint to satisfy the condition in \eqref{twistcond}. For generic choices of $a_i$ the solution preserves 2 real supercharges, i.e. it is $\tfrac{1}{16}$-BPS. When one of the $a^i$ is zero the solution is $\tfrac{1}{8}$-BPS, when two vanish $\tfrac{1}{4}$-BPS, and when three vanish $\tfrac{1}{2}$-BPS. As discussed around \eqref{eq:aglobloc} all $a_i$ consist of a global part $a^i_{\text{global}}$ and a local part $a^i_{\text{local}}$ accounting for the local contributions of the punctures. For each puncture there is a choice to add the puncture contribution to one of the four $a_i$. We can illustrate this choice by giving each puncture a colour -- red, green, blue or yellow. The puncture contributions now becomes
\begin{equation}
\begin{aligned}
&a^1_{\rm local} = \sum_{P_i = \text{green}}(1-\xi_i)\,,\quad a^2_{\rm local} = \sum_{P_i = \text{red}}(1-\xi_i)\,,\quad \\
&a^3_{\rm local} = \sum_{P_i = \text{blue}}(1-\xi_i)\,,\quad 
a^4_{\rm local} = \sum_{P_i = \text{yellow}}(1-\xi_i)\,.
\end{aligned}
\end{equation}
This construction can be phrased more geometrically in M-theory. The twisted ABJM theory describes the low-energy dynamics of $N$ M2-branes wrapped on a holomorphic two-cycle $\mathcal{C}$ in a local Calabi-Yau five-fold $X$, which is constructed as four line bundles over $\mathcal{C}$
\begin{equation}
\begin{tikzcd}
\mathbf{C}^4  \arrow[r,hook]  & \mathcal{L}_1 \oplus \mathcal{L}_2 \oplus \mathcal{L}_3 \oplus \mathcal{L}_4 \arrow[d]\\
&\mathcal{C}
\end{tikzcd}
\end{equation}
The degree of each line bundle $\mathcal{L}_i$ is $a_i$ hence the coloring indicated in which transverse direction to $\mathcal{C}$ the puncture extends. The constraint coming from the twisting, \eqref{twistcond}, translates into the Calabi-Yau condition for $X$. The local contributions $a^i_{\rm local}$ account for the local information encoded in the specific geometry of the punctures.

To fully specify the supergravity solution we need to solve also for the three scalar fields. They are expressed in terms of the flux parameters $a_i$ as
\begin{equation}
\begin{aligned}
	\e^{\lambda_1} =& \f{2(a_2+a_3)(a_1-a_4)^2-(a_1+a_4)\left( (a_2-a_3)^2 +(a_1-a_4)^2\right)-8\kappa (a_4-a_1)\mathcal{T}}{2a_4(a_4-a_1+a_2-a_3)(a_4-a_1-a_2+a_3)}\,,\\
	\e^{\lambda_2} =& \f{2(a_1+a_3)(a_2-a_4)^2-(a_2+a_4)\left( (a_1-a_3)^2 +(a_2-a_4)^2\right)-8\kappa (a_4-a_2)\mathcal{T}}{2a_4(a_4+a_1-a_2-a_3)(a_4-a_1-a_2+a_3)}\,,\\
	\e^{\lambda_3} =& \f{(a_1+a_2)(a_3-a_4)^2-(a_3+a_4)\left( (a_1-a_2)^2 +(a_3-a_4)^2\right)-8\kappa (a_4-a_3)\mathcal{T}}{2a_4(a_4+a_1-a_2-a_3)(a_4-a_1+a_2-a_3)}\,.
\end{aligned}
\end{equation}
Where we introduced the function 
\begin{equation}
	\mathcal{T} = \f12\left( 1-3\sum_i (a_i)^2 \right)^{1/2}-8 \sqrt{a_1 a_2 a_3 a_4}\,.
\end{equation}
Finally the constants appearing in the metric are given by
\begin{equation}
\begin{aligned}
	\e^{2f_0} =& \f{2\e^{\f12(\lambda_1+\lambda_2+\lambda_3)}}{ \left(1+\e^{\lambda_1}+\e^{\lambda_2}+\e^{\lambda_3}\right)^2} \,,\\
	\e^{\varphi_0} =& \f{1}{2} \e^{-\f12(\lambda_1+\lambda_2+\lambda_3)}\left(a_1 \e^{\lambda_2+\lambda_3}+a_2 \e^{\lambda_1+\lambda_3}+a_3 \e^{\lambda_1+\lambda_2}-a_4 \e^{\lambda_1+\lambda_2+\lambda_3}\right) \,.
\end{aligned}
\end{equation}
We can uplift these solutions to solutions of eleven-dimensional supergravity using the uplift formulae in Appendix \ref{app:upliftM2}. Locally around each of the punctures we can analyze the uplifted solution. The puncture locally preserve half of the maximal supersymmetry so we can make a local gauge and coordinate transformation such that $a_1=\chi_{\mathbf{g},\xi}$ and $a_2=a_3=a_4=0$ and $\xi=\f{1}{k}$. In that case we find that the eleven-dimensional solution is regular up to a $\mathbf{Z}_k$ singularity at $\alpha = 0$. Near that point the uplifted metric becomes
\begin{equation}
\dd s_{11}^2 = \Delta^{1/2}\left[ \dd s_{\text{AdS}_2}^2 + \dd s^2_{S^5} + \dd s^2_{\mathbf{R}^4/\mathbf{Z}_k} \right]\,,
\end{equation}
which matches the expectation \eqref{Zksing} from the general discussion above.

\subsection{D3-branes on singular curves}

The gauge theory living on the worldvolume of $N$ D3-branes is given by four-dimensional $\mathcal{N}=4$ super Yang-Mills theory which at large $N$ and large 't Hooft coupling is dual to ten-dimensional type IIB supergravity on $\text{AdS}_5\times S^5$. Compactifying the $\mathcal{N}=4$ theory on a the surface $\mathcal{C}$ is described holographically by a ten-dimensional supergravity background which is asymptotically locally $\text{AdS}_5\times S^5$ but for which the topology at a fixed value of the radial coordinate is an $S_5$ fibration over $\mathcal{C}$ \cite{Maldacena:2000mw,Benini:2013cda}. Once again the construction of these solutions is most efficient in a truncation of the maximal five-dimensional $\SO(6)$ gauged supergravity \cite{Gunaydin:1984qu,Pernici:1985ju,Gunaydin:1985cu} studied in \cite{Cvetic:1999xp}. This truncation contains the metric, three Abelian gauge fields in the Cartan of the $\SO(6)$ gauge group and two real scalars. All solutions of this truncated theory can be uplifted to solutions of type IIB supergravity on $S^5$ \cite{Cvetic:1999xp}. 

In \cite{Benini:2013cda} the near-horizon geometry of $N$ D3-branes wrapped on a smooth Riemann surface was analyzed using this truncation. We can extend this analysis by using the more general Ansatz in \eqref{Ansatz}. The BPS equations then reduce to the Liouville equation \eqref{Liouville} for the conformal factor $\varphi$. In terms of this conformal factor, the field strengths are given by
\begin{equation}
F^{(i)} = \kappa\f{a_i\pi}{g} \vol^{-1}_{\mathbf{g},\xi} \, \dvol_{\mathbf{g},\xi}\,,
\end{equation}
where $g$ is the gauge coupling of the supergravity theory and $i=1,2,3$. The constants $a_i$ determine the choice of topological twist and have to satisfy \eqref{twistcond}. For generic choices of $a^i$ the theory preserves $\mathcal{N}=(0,2)$ supersymmetry, when one of the $a^i$ is zero and the other two are equal we get  $\mathcal{N}=(2,2)$, when two $a^i$ vanish the supersymmetry is $\mathcal{N}=(4,4)$ and when all $a^i$ vanish (and $\mathbf{g}=1$) we have $\mathcal{N}=(8,8)$ supersymmetry. As in \eqref{eq:aglobloc} all $a_i$ consist of a regular part $a^i_{\text{global}}$ and a local part associated to the punctures $a^i_{\text{local}}$. For each puncture we have the choice to add the puncture contribution to one of the three $a_i$. We represent this choice by assigning a color to each puncture -- red, green or blue. Then the puncture contribution to the $a_i$ becomes
\begin{equation}
	a^1_{\rm local} = \sum_{P_i = \text{green}}(1-\xi_i)\,,\qquad a^2_{\rm local} = \sum_{P_i = \text{red}}(1-\xi_i)\,,\qquad a^3_{\rm local} = \sum_{P_i = \text{blue}}(1-\xi_i)\,,\qquad 
\end{equation}
The geometric interpretation of this construction is by now familiar The twisted $\mathcal{N}=4$ theory describes the low-energy dynamics of $N$ D3-branes wrapped on a holomorphic two-cycle $\mathcal{C}$ in a local Calabi-Yau four-fold X, which is composed of three line bundles over $\mathcal{C}$
\begin{equation}
\begin{tikzcd}
\mathbf{C}^3  \arrow[r,hook]  & \mathcal{L}_1\oplus \mathcal{L}_2\oplus \mathcal{L}_3 \arrow[d]\\
&\mathcal{C}
\end{tikzcd}
\end{equation}
As before, the degree of each line bundle $\mathcal{L}_1$ is $a_i$ and the coloring describes in which part of the line bundle the puncture. Finally the twist condition \eqref{twistcond} translates into the Calabi-Yau condition on $X$. The local information of the puncture is captured by $a^i_{\rm local}$. 

The solution for the two scalars, $\lambda_1$ and $\lambda_2$, is given by
\begin{equation}
\begin{aligned}
\e^{3\lambda_1+\lambda_2} =& \frac{a_3(a_1+ a_2-a_3)}{a_1(-a_1+a_2+a_3)}  \,,\\
\e^{2\lambda_2} =& \frac{a_2(a_1- a_2+a_3)}{a_1(-a_1+a_2+a_3)}  \,,\\
\end{aligned}
\end{equation}
and the constants appearing in the metric by
\begin{equation}
\begin{aligned}
\e^{3f_0} =& \f{a_1a_2(a_1- a_2-a_3)(a_1+a_2-a_3)(a_1- a_2+a_3)}{(a_1^2+a_2^2+a_3^2-2(a_1a_2+a_1a_3+a_2a_3))^3}\,,\\
\e^{3\varphi_0} =& \f{1}{2^6} \f{a_1^{2}a_2^{2}a_3^{2}}{(a_1+ a_2-a_3)(a_1- a_2+a_3)(-a_1+ a_2+a_3)}\,.
\end{aligned}
\end{equation}
We can uplift these five-dimensional solutions to type IIB supergravity using the uplift formulae in Appendix \ref{app:upliftD3} and analyzethe solution  locally  around each puncture. The punctures locally preserve $\mathcal{N}=(4,4)$ supersymmetry so we can make a local change of coordinates such that $a_1=\chi_{\mathbf{g},\xi}$ and $a_2=a_3=0$ and $\xi=\f{1}{k}$. The result is a regular solution up to a $\mathbf{Z}_k$ singularity at $\alpha = 0$. Near this point the uplifted metric becomes
\begin{equation}
	\dd s_{10}^2 = \Delta^{1/2}\left[ \dd s_{\text{AdS}_3}^2 + \dd s^2_{S^3} + \dd s^2_{\mathbf{R}^4/\mathbf{Z}_k} \right]
\end{equation}
which matches the expectation \eqref{Zksing} from the general discussion above.

\subsection{D4-D8-branes on singular curves}

The gauge theory living on the worldvolume of a stack of $N$ D4-branes in a background of $N_f$ D8-branes with an O8-plane is non-renormalizable but flows to a five-dimensional $\mathcal{N}=1$ SCFT in the UV. At large $N$ the SCFT is dual to ten-dimensional massive type IIA supergravity on AdS$_6\times S^4/\mathbf{Z}_2$ \cite{Brandhuber:1999np}. A twisted compactification of the SCFT on a curve $\mathcal{C}$ results in a ten-dimensional supergravity background which is asymptotically locally $\text{AdS}_6\times S^4/\mathbf{Z}_2$ but for which the topology at a fixed value of the radial coordinate is an $S^4/\mathbf{Z}_2$ fibration over $\mathcal{C}$, see \cite{Nunez:2001pt,Bah:2018lyv,Crichigno:2018adf}. Massive type IIA supergravity admits a truncation to the Romans six-dimensional $\SU(2)$ gauged supergravity \cite{Romans:1985tw}. For the solutions of interest we can restrict to a further truncation containing only the metric, an Abelian gauge field in the Cartan of the $\SU(2)$ gauge group and one real scalar. All solutions of this truncation can be uplifted to solutions of massive type IIA supergravity \cite{Cvetic:1999un}.\footnote{Solutions of the six-dimensional Romans supergravity can also be uplifted to type IIB supergravity. We do not consider this possibility here and restrict to uplifts to type IIA.}

As shown in \cite{BGP} inserting the Ansatz \eqref{Ansatz} in the BPS equations of the six-dimensional supergravity  leads to the Liouville equation for $\varphi$. Since the supergravity truncation has only one gauge field there is only one possible twist and all fields are fully fixed in terms of $\varphi$. The field strength $F$ is given by
\begin{equation}
F = -\f{\kappa}{2 g}\dvol_{\mathbf{g},\xi}\,.
\end{equation}
The scalar $\lambda$ and the metric constants are given by
\begin{equation}
\e^{4\lambda} = \f23  \,,\qquad \e^{f_0} = \f{3^{1/4}}{2^{3/4}} \e^{-\lambda}\,,\qquad \e^{\varphi_0} = \f{\sqrt{3}}{8\sqrt{2}}\,.
\end{equation}
Here $g$ is the gauge coupling of the six-dimensional supergravity which is related to the mass parameter $m$ of massive type IIA supergravity by $m=\f{\sqrt{2}}{3}g$.

We can uplift this solution to massive type IIA supergravity using the uplift formulae of \cite{Cvetic:1999un} (which are summarized in Appendix \ref{app:upliftD4D8}) and study the resulting ten-dimensional solution near a conical defect with opening angle $\xi$. Due to the fact that the uplift only includes half of the four-sphere, $S^4/\mathbf{Z}_2$ this case deviates slightly from the general story. The resulting geometry takes the form
\begin{equation}
	\dd s_{10}^2 \simeq \Delta^{3/8}\left[ \dd s^2_{\text{AdS}_4}  + \dd s^2_6 \right] \,,
\end{equation}
where the six-dimensional internal space is given by
\begin{equation}
	\dd s_6^2 = \frac{3}{8}\dd \alpha^2 + \xi^2\rho^2\dd\theta^2 + \dd \rho^2 + (\sigma_1^2+\sigma_2^2) + (\sigma_3+(1-\xi)\dd\theta)^2\,,
\end{equation}
and the Riemann surface is parametrized by the coordinates $\rho$ and $\theta$. The $(\sigma_1^2+\sigma_2^2)$ part of the metric has the correct symmetries to account for the $\U(1)_R$ R-symmetry of a three-dimensional $\mathcal{N}=2$ theory however it is non-trivially fibered over the remainder of the internal space. Analyzing the global structure of this solution goes beyond the scope of this work.

\subsection{M5-branes on singular curves}
\label{subsec:M5punc}

The gauge theory living on the worldvolume of $N$ M5-branes is the six-dimensional $\mathcal{N}=(2,0)$ theory of type $A_{N-1}$. At large $N$ this theory is dual to eleven-dimensional supergravity on $AdS_7\times S^4$. The large $N$ dual of a twisted compactification of this theory on a complex curve $\mathcal{C}$ is an eleven-dimensional supergravity background which is asymptotically locally $AdS_7\times S^4$ but for which the topology at a fixed value of the radial coordinate is an $S^4$ fibration over $\mathcal{C}$. Eleven-dimensional supergravity admits a consistent truncation to the lowest Kaluza-Klein modes given by maximal $\SO(5)$ gauged supergravity in seven dimensions \cite{Pernici:1984xx}. Once more we can restrict to a further truncation of the theory, containing only the metric, two abelian gaage fields in the Cartan of the $\SO(5)$ gauge group, and two real scalars parametrizing the squashing of the $S^4$ \cite{Cvetic:1999xp,Liu:1999ai}. Moreover, all solutions we obtain in seven dimensions can be uplifted to eleven dimensions using the results in \cite{Nastase:1999kf,Nastase:1999cb,Cvetic:1999xp}.

The near-horizon geometry of the M5-branes wrapping smooth curves was considered in \cite{Maldacena:2000mw,Bah:2012dg}. We summarize the derivation of the BPS equations for this construction in Appendix \ref{app:BPS}. Yet again these equations reduce to the Liouville equation for $\varphi$ and  all other fields are determined in terms of this function only. The field strengths are given by
\begin{equation}\label{eq:F1F2M5}
\begin{aligned}
F^{(1)} =& -\kappa\frac{a_1 \pi}{4g}\vol_{\mathbf{g},\xi}^{-1} \, \dvol_{\mathbf{g},\xi}\,,\\
F^{(2)} =& -\kappa\frac{a_2\pi}{4g}\vol_{\mathbf{g},\xi}^{-1}\, \dvol_{\mathbf{g},\xi}\,,
\end{aligned}
\end{equation}
where $g$ is the supergravity gauge coupling and we have the usual condition \eqref{twistcond}. As in the previous cases we can assign a color to each puncture -- red or blue -- indicating to which of the $a_i$ it contributes,
\begin{equation}
a^1_{\rm local} = \sum_{P_i = \text{red}}(1-\xi_i)\,,\qquad a^2_{\rm local} = \sum_{P_i = \text{blue}}(1-\xi_i)\,. 
\end{equation}
This construction can again be interpreted geometrically as the low energy dynamics of $N$ M5-branes wrapping a holomorphic two-cycle $\mathcal{C}$ in a Calabi-Yau three-fold $X$ with local geometry
\begin{equation}\label{eq:M5CYcurve}
\begin{tikzcd}
\mathbf{C}^2  \arrow[r,hook]  & \mathcal{L}_1\oplus \mathcal{L}_2 \arrow[d]\\
&\mathcal{C}
\end{tikzcd}
\end{equation}
Where $\mathcal{L}_1$ and $\mathcal{L}_2$ are two line bundles of degree $a_i$ and the twisting condition for the $a_i$ again translates into the Calabi-Yau condition for $X$. 

The solution for the supergravity scalars $\lambda_1$ and $\lambda_2$ is given by
\begin{equation}\label{eq:lam12M5sol}
\begin{aligned}
\e^{10\lambda_1} =& \frac{1+7z +7z^2 + 33z^3+\kappa(1+4z+19z^2)\sqrt{1+3z^2}}{4z(1-z)^2}  \,,\\
\e^{2(\lambda_1-\lambda_2)} =& \f{2z-\kappa\sqrt{1+3z^2}}{1+z}  \,.\\
\end{aligned}
\end{equation}
where as in \cite{Bah:2011vv,Bah:2012dg} we have defined
\begin{equation}\label{eq:zdef}
z = \frac{a_1-a_2}{a_1+a_2}\,.
\end{equation}
Finally the constants appearing in the metric are given by
\begin{equation}\label{eq:ef0eg0M5}
\begin{aligned}
\e^{f_0} =& \f12 \e^{4\lambda_1+4\lambda_2}\,,\\
\e^{\varphi_0} =&  \f{\e^{2\lambda_1+2\lambda_2}}{64}\left( (1+z)\e^{2\lambda_2} + (1-z)\e^{2\lambda_1}  \right)\,.
\end{aligned}
\end{equation}
Once more, we can analyse the uplifted geometry locally around a puncture with opening angle $\xi=\f{1}{k}$ using the uplift formulae of \cite{Cvetic:1999xp}, summariz ed in Appendix \ref{app:upliftM5}. Locally around the puncture we preserve $\mathcal{N}=2$ supersymmetry and we can thus, without loss of generality, restrict to the case $a_2=0$. The uplifted solution gives a regular geometry up to a single $\mathbf{Z}_k$ singularity at $\alpha = 0$. The geometry near this singularity takes the form
\begin{equation}
\dd s_{11}^2 = \Delta^{1/2}\left[ \dd s_{\text{AdS}_5}^2 + \dd s^2_{S^2} + \dd s^2_{\mathbf{R}^4/\mathbf{Z}_k} \right]
\end{equation}
which again is in line with the general discussion above.

\section{Wrapped M5-branes and twists of the $(2,0)$ theory} 
\label{sec:M5branes}

From now on we focus on M5-branes and will accumulate evidence for the claim that punctures can indeed be treated in gauged supergravity as discussed in previous section. We start by reviewing the six-dimensional $\mathcal{N}=(2,0)$ theory and its partial topological twists and discuss their realization as the worldvolume theory of M5-branes wrapped on complex curves. For concreteness here and in most of the following we limit ourselves to $\mathcal{N}=(2,0)$ theories of type $A_{N-1}$. Parts of our analysis admits generalizations to $D_N$ or $E_{6,7,8}$ type theories.

\subsection{Partial twists of the $\mathcal{N}=(2,0)$ theory}

We are interested in the six-dimensional $\mathcal{N}=(2,0)$ theory of type $A_{N-1}$ defined on a spacetime of the form
\begin{equation}\label{}
	\mathbf{R}^{1,3}\times \mathcal{C}\,,
\end{equation}
where $\mathcal{C}$ is a Riemann surface of genus $\mathbf{g}>1$ with prescribed singularities. To preserve supersymmetry  on $\mathbf{R}^{1,3}$ we perform a partial topological twist \cite{Witten:1988ze,Bershadsky:1995qy} by turning on a background flux for the $\SO(5)$ R-symmetry of the $(2,0)$ theory. A choice of twist corresponds to a choice of abelian subgroup $\U(1)^\prime \subset \U(1)_\mathcal{C}\times \SO(5)_R$ such that a number of supercharges are invariant under $\U(1)^\prime$. Here $\U(1)_\mathcal{C}$ is the structure group of the Riemann surface.

Since only an abelian factor of the structure group is being twisted it suffices to look at the Cartan of the $R$-symmetry group, i.e. $\U(1)_+\times \U(1)_- \subset \SO(5)_R$. Under the subgroup $\SO(1,3)\times \U(1)_\mathcal{C}\times \U(1)_+\times \U(1)_- \subset \SO(1,5)\times \SO(5)_R$, the supercharges of the $(2,0)$ theory decompose as
\small
\begin{equation}
\mathbf{4}\times \mathbf{4} \rightarrow \left[ \left( \mathbf{2},\mathbf{1} \right)_{\f12}\oplus \left( \mathbf{1},\mathbf{2} \right)_{-\f12} \right]\otimes \left[ \left(\mathbf{\f12},\mathbf{\f12}\right)\oplus \left(-\mathbf{\f12},\mathbf{\f12}\right)\oplus \left(\mathbf{\f12},-\mathbf{\f12}\right)\oplus \left(-\mathbf{\f12},-\mathbf{\f12}\right) \right]\,,
\end{equation}
\normalsize
and satisfy a reality constraint coming from the symplectic-Majorana condition. Thus under the $\U(1)$ subgroup generated by a Lie algebra element $\mathfrak{t}^\prime = \mathfrak{t}_\mathcal{C} + a\, \mathfrak{t}_+ + b\, \mathfrak{t}_-$, where the $\mathfrak{t}$'s are the generators of the respective $\U(1)$'s, the supercharges transform with charges $\pm \f12 \pm \f{a}{2} \pm \f{b}{2}$. For any choice of $a$ and $b$ such that $a\pm b = \pm 1$ there are at least four real supercharges. Choosing $a=\f{a_1}{a_1+a_2}$ and $b=\frac{a_2}{a_1+a_2}$ we can identify the holonomy group $\U(1)_h$ as the linear combination
\begin{equation}
\U(1)_h = \f{a^1}{a^1+a^2}\U(1)_+ + \f{a^2}{a^1+a^2}\U(1)_-\,.
\end{equation}
This twist in general preserves four supercharges and therefore leads to an $\mathcal{N}=1$ supersymmetric field theory in four dimensions. The field theory has $\U(1)^2$ flavour symmetry with generators
\begin{equation}
R_0 = \f12(\mathfrak{t}_+ + \mathfrak{t}_-)\,, \qquad \mathcal{F} = \f12 (\mathfrak{t}_+-\mathfrak{t}_-)\,.
\end{equation}
Where $R_0$ is an $R$-symmetry. In the IR the $\mathcal{N}=1$ superconformal $R$-symmetry will in general be given by a combination
\begin{equation}
R_{\mathcal{N}=1} = R_0 + \epsilon \mathcal{F}\,.
\end{equation}
of the two $\U(1)$s. The value of $\epsilon$ is a priori unknown but will be fixed by $a$ maximization \cite{Intriligator:2003jj}.

When either $a_1$ or $a_2$ vanishes, the theory preserves eight supercharges and either $\U(1)_\pm$ is enhanced to an $\SU(2)_\pm$ R-symmetry. In this case twisted compactifications of the $\mathcal{N}=(2, 0)$ theory flow to four-dimensional $\mathcal{N}=2$ SCFTs of class $\mathcal{S}$ \cite{Gaiotto:2009we}. When $a_1=a_2\neq0$ the diagonal subgroup $\U(1)\subset \U(1)_+\times \U(1)_-$ is used to perform the twist. This procedure preserves the diagonal subgroup $\SU(2)_\mathcal{F}\subset \SU(2)_+\times \SU(2)_-$ and consequently the flavour symmetry is enhanced from $\U(1)_\mathcal{F}$ to $\SU(2)_\mathcal{F}$. This corresponds to the class $\mathcal{N}=1$ SCFTs studied in \cite{Maldacena:2000mw,Benini:2009mz}.

In M-theory we can construct these partially twisted theories by wrapping M5-branes on a complex curve with prescribed singularities. We can decompose the eleven-dimensional spacetime as
\begin{equation}
	M^{1,10} \rightarrow \mathbf{R}^{1,3}\times \mathbf{R} \times CY_3\,.
\end{equation}
The M5-branes extend along $\mathbf{R}^{1,3}$ and wrap a complex curve $\mathcal{C}\subset CY_3$ inside the Calabi-Yau threefold. In general this Calabi-Yau threefold is an $\SU(2)$ bundle over the curve whose determinant line bundle equals the canonical line bundle $K_\mathcal{C}$ of the curve. When the structure group is reduced from $\SU(2)$ to $\U(1)$, in addition to the $\U(1)_R$ R-symmetry, the local geometry enjoys an additional $\U(1)_{\mathcal{F}}$ flavor symmetry under which the supercharges are invariant. Under these circumstances the local geometry takes the form presented in \eqref{eq:M5CYcurve} where $\mathcal{L}_1$ and $\mathcal{L}_2$ are two complex line bundles subject to the condition $\mathcal{L}_1 \otimes \mathcal{L}_2 = K_\mathcal{C}$. While the Chern class fails to be well-defined for singular Calabi-Yau's, the canonical bundle and canonical class can still be defined for mild singularities.\footnote{The criterion for the singularities to be mild enough to still be able to define the canonical class is that all singularities have to be Gorenstein, this is the case for all singularities we consider.} The two line bundles are associated to the $\U(1)_\pm$ above and the Calabi-Yau condition simply reproduces the twist condition
\begin{equation}\label{CYcondition}
a_1+a_2 = -\chi(\mathcal{C},\beta) = 2\mathbf{g}-2+\sum_{j=1}^n\beta_j\,,
\end{equation}
where $n$ is the number of singularities, $\beta_j$ is a puncture dependent contribution, $\chi(\mathcal{C},\beta)$ is the modified Euler characteristic of the singular curve and $a^1$ and $a^2$ are the degrees of the line bundles, 
\begin{equation}\label{ChernNum}
	\deg(\mathcal{L}_1) = a_1\,,\qquad \deg(\mathcal{L}_2) = a_2\,.
\end{equation}
For conical singularities on the Riemann surface the puncture contribution is given exactly by $\beta_j = 1-\xi_j$ where $\xi_j$ is defined in Section~\ref{sec:puncbranes} as the defect angle at the conical singularity. For different choices of $a_1$ and $a_2$ the fields of the M5-branes transform in different representations of the flavor symmetry $\U(1)_\mathcal{F}$ and one generically ends up in different $\mathcal{N}=1$ IR fixed points.

\subsection{Central charges from the anomaly polynomial}

A powerful tool to study the six-dimensional $\mathcal{N}=(2,0)$ theory and its partial topological twists is provided by anomalies. The central charges of the resulting four-dimensional theory can be computed by integrating the M5-brane anomaly polynomial over the curve $\mathcal{C}$. This procedure was introduced in \cite{Alday:2009qq,Chacaltana:2012zy,Bah:2018gwc} and further explored in the present context in \cite{Bah:2018gwc,Bah:2018jrv,Bah:2019jts,Bah:2019rgq}. Here we summarize the main ingredients of these calculations.

The $a$ and $c$ anomaly of a four-dimensional $\mathcal{N}=1$ SCFT are completely determined by the linear and cubic 't Hooft anomalies of the superconformal R-symmetry \cite{Anselmi:1997am},
\begin{equation}\label{acRanom}
	a = \f{3}{32}\left(3\Tr R_{\mathcal{N}=1}^3 - \Tr R_{\mathcal{N}=1} \right) \,, \qquad c=\f{1}{32} \left(9\Tr R_{\mathcal{N}=1}^3-5\Tr R_{\mathcal{N}=1}\right)\,.
\end{equation}
These anomalies can be read off from the six-form anomaly polynomial given by
\begin{equation}\label{an6}
	I_6 = \f{\Tr R_{\mathcal{N}=1}^3}{6}c_1(F)^3 - \f{\Tr R_{\mathcal{N}=1}}{24}c_1(F)p_1(T_4)\,,
\end{equation}
where $F$ is the $\U(1)$ bundle which couples to the R symmetry and $T_4$ is the tangent bundle to the four-dimensional spacetime manifold. This anomaly six-form can in turn be obtained by integrating the anomaly eight-form of the six-dimensional theory over a (possibly) singular curve $\mathcal{C}$. As explained in \cite{Bah:2018gwc} the contribution to the anomaly six-form can be separated in two parts, one geometric part and a second part accommodating the local contributions of the different singularities on $\mathcal{C}$
\begin{equation}
	I_6 = I_6(\mathcal{C}) + \sum_{i} I_6 (P_i)\,. 
\end{equation}
We discuss these two contributions separately below.

\subsubsection{Bulk contribution}

We start by computing the geometric part of the anomaly polynomial. In our treatment we will apply a slightly different split than the one in \cite{Bah:2018gwc} by defining the geometric part to contain information only about the smooth Riemann surfaces. All information about the punctures is packaged in the local contributions. The anomaly eight-form for a single M5-brane is given by \cite{Witten:1996hc,Harvey:1998bx,Intriligator:2000eq}
\begin{equation}
	I_8[1] = \f{1}{48}\left[ p_2\left(NW\right)-p_2\left(TW\right) + \f14 \left(p_1\left(TW\right)-p_1\left( NW \right)\right)^2 \right]\,,
\end{equation}
where by $NW$ and $TW$ we denote the normal and tangent bundle to the brane world volume and $p_1$ and $p_2$ are the first and second Pontryagin classes. For a general $\mathcal{N}=(2,0)$ theory of type $\mathfrak{g}\in $ ADE, the anomaly polynomial takes the form
\begin{equation}\label{an8}
	I_8[\mathfrak{g}] = r_G I_8[1]  + \f{d_G h_G}{24}p_2\left(NW\right)\,.
\end{equation} 
Here $r_G$, $d_G$, and $h_G$ stand for the rank, dimension and Coxeter number of the group $G$, see Table~\ref{tab:rdh}. The normal bundle can be thought of as the $\SO(5)$ bundle coupled to the R-symmetry of the six-dimensional theory.
\begin{table}[H]
	\centering
	\setlength{\tabcolsep}{20pt}
	\begin{tabular}{c|ccc}
		$\mathfrak{g}$ & $r_\mathfrak{g}$ & $d_\mathfrak{g}$ & $h_\mathfrak{g}$\\
		\hline\rule{0pt}{2.6ex}
		$A_{N-1}$ & $N-1$ & $N^2-1$ & $N$ \\
		$D_{N}$ & $N$ & $N(2N-1)$ & $2N-2$ \\
		$E_{6}$ & $6$ & $78$ & $12$ \\
		$E_{7}$ & $7$ & $133$ & $18$ \\
		$E_{8}$ & $8$ & $248$ & $30$ \\
	\end{tabular}
	\caption{Rank, dimension and Coxeter number for the simply laced Lie algebras.}
	\label{tab:rdh}
\end{table}
The first and second Pontryagin classes of a vector bundle $E$ can be expressed in terms of the Chern roots $e_i$ as 
\begin{equation}
p_1(E) = \sum_{i}e_i^2\,,\qquad p_2(E) = \sum_{i<j}e_i^2 e_j^2\,.
\end{equation}
To compute the anomaly six-form for a $\U(1)$ R-symmetry of the form
\begin{equation}
	R_{\mathcal{N}=1} = R_0 + \epsilon \mathcal{F} \,,
\end{equation}
we need to couple the symmetry to a non-trivial $\U(1)$ bundle $\mathcal{F}$ over the flat four-dimensional part of the brane worldvolume. This induces a shift in the Chern classes
\begin{equation}
	c_1(\mathcal{L}_1)\rightarrow  c_1(\mathcal{L}_1) + (1+\epsilon) c_1(\mathcal{F})\,,\qquad 	c_1(\mathcal{L}_2)\rightarrow  c_1(\mathcal{L}_2) + (1-\epsilon) c_1(\mathcal{F})\,.
\end{equation}
There are an infinite number of such decomposable bundles over the smooth Riemann surface, labelled by the Chern numbers of the line bundles
\begin{equation}
	c_1(\mathcal{L}_1) = p\,,\qquad c_1(\mathcal{L}_2) = q\,.
\end{equation}
The Calabi-Yau condition in this case reduces to $p+q = 2g-2$. We can now integrate the eight-form \eqref{an8} over the smooth curve $\mathcal{C}$ to obtain
\begin{multline}
	I_6(\mathcal{C}) = \int_{\mathcal{C}}I_8[\mathfrak{g}] = -\f{\chi(\mathcal{C})}{12}\Big[(r_G+d_Gh_G)(1+\mathfrak{z} \epsilon^3) - d_Gh_G (\epsilon^2+\mathfrak{z} \epsilon)\Big]c_1(\mathcal{F})^3 \\
	+\f{\chi(\mathcal{C})}{24}r_G(1+\mathfrak{z}\epsilon)c_1(\mathcal{F})p_1(T_4)\,,
\end{multline}
where $\chi(\mathcal{C})$ is the Euler characteristic of the curve $\mathcal{C}$ and we have defined
\begin{equation}\label{eq:frakzdef}
	\mathfrak{z} = \frac{p-q}{p+q}\,.
\end{equation}
%

\subsubsection{Punctured intermezzo}

\begin{figure}[H]
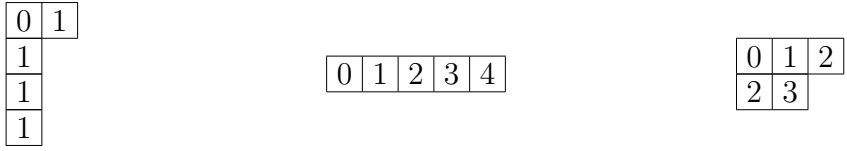

	\centering
	\begin{minipage}{0.32\textwidth}
		\centering
		\young(01,1,1,1)
	\end{minipage}
	\begin{minipage}{0.32\textwidth}
		\centering
		\young(01234)
	\end{minipage}
	\begin{minipage}{0.32\textwidth}
		\centering
		\young(012,23)
	\end{minipage} 
	\caption{Labelling Young tableaux by their pole structure $p_k$. The first Young tableau corresponds to the minimal puncture with $p= (0,1,1,1,1)$. The second one is the maximal puncture with $p=(0,1,2,3,4)$ and the third diagram corresponds to an intermediate puncture with $p = (0,1,2,2,3)$.}
	\label{fig:Younglab}
\end{figure}

To compute the local contributions from each puncture to the anomaly polynomial we need some more information about the different types of punctures that can appear in our setup. We only consider punctures that locally preserve $\mathcal{N}=2$ supersymmetry. Each such puncture is characterized by an embedding $\rho: \su(2)\rightarrow \mathfrak{g}$ and a sign $\sigma_i = \pm 1$. The flavor symmetry of the puncture is determined as the commutant $\mathfrak{h}\subset\mathfrak{g}$ of the image of $\rho$ and the sign determines whether the puncture preserves the $\U(1)_+ \times \SU(2)_-$ or $\SU(2)_+ \times \U(1)_-$ symmetry. This $\mathbf{Z}_2$ valued label determines the directions normal to the M5-branes along which the puncture extends. We represent this label by coloring each puncture as in in Figure~\ref{fig:congen}.
For $\mathfrak{g}=A_{N-1}$, the choice of embedding $\rho$ is in one-to-one correspondence with a partition of $N$ and thus with a Young tableau $Y$. A Young tableau with $n_h$ columns of length $h$ corresponds to a puncture $P$ with global flavour symmetry group 
\begin{equation}
	G_{P} = \text{S}\left(\prod_{h}\U(n_h)\right)\,.
\end{equation}
A maximal puncture is represented by a Young tableau with a single row of length $N$, see Figure \ref{fig:Younglab}, and has the maximal amount, i.e. $\SU(N)$, of global symmetry associated to it, a minimal (or simple) puncture is represented by a Young tableau with one row of length 2 and $N-2$ rows of length $1$ and preserves the minimum amount of global symmetry, namely $\U(1)$. Equivalently one can label every puncture with a set of integers, $p_k$, characterizing the pole structure of the degree $k$ Seiberg-Witten differentials at the puncture. These integers can be obtained from the Young tableaux as follows: Start with the first row and label the first box with $p_1 =0$, then increase the label by one as you move to the right along the first row of the Young tableau. When this row is finished, move to the next row and label the first box of the row with the same label as the last box in the previous row. Repeat this labelling process in every row until all $N$ boxes have a label $p_k$. This labelling procedure is illustrated in Figure~\ref{fig:Younglab} for some simple examples with $N=5$.
To each puncture one can associate an effective number of vector multiplets, $n_v(P_i)$, and hypermultiplets, $n_h(P_i)$, given by \cite{Gaiotto:2009we,Chacaltana:2010ks}
\begin{equation}
	n_v(P_i) = \sum_{k=2}^{N}(2k-1)p_k\,,\qquad n_h(P_i) = n_v(P_i) + \f12\left(-(1+r_G) + \sum_{r}l_r^2\right)\,,
\end{equation}
where $l_r$ is the length of the $r$th row of the Young tableau. The numbers $n_v$ and $n_h$ represent the effective degrees of freedom of the specific puncture\footnote{The definition we use differs slightly from the ones in \cite{Chacaltana:2010ks}. Our definition only accounts for the local degrees of freedom near the punctures, all global information is absorbed in $I_6(\mathcal{C})$.} and indeed when considering free theories these numbers agree with the actual number of vector and hypermultiplets. These constants can now be used to determine the local contribution of each puncture to the anomaly polynomial.

\subsubsection{Contribution from a puncture}

After this short intermezzo, we are ready to compute the local contribution of each puncture $P_i$ to the anomaly polynomial. A puncture (locally) preserving flavor symmetry $G$ with $\mathbf{Z}_2$-label $\sigma_i$ contributes the following \cite{Tachikawa:2015bga,Bah:2018gwc}:
\begin{multline}
I_6(P_i) = \f{1}{6}\left( (1+\sigma_i \epsilon^3)n_v(P_i)-\f14(1+\sigma_i\epsilon)^3 n_v(P_i)\right)c_1(\mathcal{F})^3 \\
+\f{1}{24}\left(1+\sigma_i\epsilon\right)\left(n_h(P_i)-n_v(P_i)\right)c_1(\mathcal{F})p_1(T_4)\\
 - \f{k_G}{3}\left(1+\sigma_i\epsilon\right)c_1(\mathcal{F})c_2(F_G) \,.
\end{multline}
For a puncture associated to a Young tableau $Y$, the central charge of the flavor symmetry factor $SU(n_h)$ is given by
\begin{equation}
         k_{\SU(n_h)} = 2 \sum_{i\leq h} s_i\,,
\end{equation}
where $s_i$ is the length of the $i^{\rm th}$ row of $Y^T$, the transpose of the original Young tableau.

Summing over all punctures in the theory one can easily read off the 't Hooft anomalies of the four-dimensional R-symmetry and compute the trial central charges as a function of $\epsilon$. The correct value of the central charge is then obtained by maximizing $a$ with respect to $\epsilon$. When we set all $\sigma=1$ and $q=0$ (or $\sigma=-1$ and $p=0$) one finds an $\mathcal{N}=2$ theory and one can check that the anomalies reduces to the known results for $\mathcal{N}=2$ class $\mathcal{S}$ theories \cite{Gaiotto:2009gz}. In the limit with no punctures the anomalies reduce to the result obtained in \cite{Bah:2012dg}.

\section{Punctures in gauged supergravity} 
\label{sec:punctures}

In this section we come back to the solution described in Section \ref{subsec:M5punc} and carefully study the geometry around a puncture, first in eleven and then in seven dimensions. Once we have understood how to describe punctures, we compute the holographic central charges of these AdS$_5$ solutions for a generic punctured Riemann surface and find an exact match at leading order in $N$ with the results in Section~\ref{sec:M5branes}. Furthermore, we compute the dimension of a protected operator corresponding to an M2-branes wrapped on the Riemann surface and discuss the exactly marginal deformations of our solutions.

\subsection{Seven-dimensional supergravity solution}
\label{subsec:7dM5sol}

The seven-dimensional solution corresponding to M5-branes wrapped around a curve $\mathcal{C}$ takes the form \eqref{Ansatz}
\begin{equation}
\begin{aligned}
\dd s^2 =& \frac{4\e^{2f_0}}{g^2r^2}(-\dd t^2 + \dd z_1^2+ \dd z_2^2+ \dd z_3^2 + \dd r^2) + \f{4\e^{\varphi_0+\varphi(x_1,x_2)}}{g^2} (\dd x_1^2 + \dd x_2^2)\,,\\
A^{(i)} =& A^{(i)}_{x_1}(x_1,x_2)\dd x_1+ A^{(i)}_{x_2}(x_1,x_2)\dd x_2\,.
\end{aligned}
\end{equation}
where $g$ is the coupling constant of the gauged supergravity, $i= 1,2$ and the scalars $\lambda_i$ take constant values in the IR. The Riemann surface has Gaussian curvature $\kappa=\pm 1 , 0$. In this section we consider only $\kappa = \pm1$. The analysis for $\kappa=0$, i.e. the torus without punctures or the flat punctured sphere\footnote{A sphere with $n$ punctures with puncture contributions $\sum_{i=1}^{n}(1-\xi_i)=2$ summing exactly to 2}, deviates slightly from the discussion below. For the torus without punctures our solution reduces to the one found in \cite{Bah:2012dg}. For the flat sphere the discussion is very similar but since the Liouville equation \eqref{Liouville} reduces to the Laplace equation one has to study the solution of this equation with singular sources. We will not investigate this situation in detail here.

As discussed above the BPS equations governing this setup reduce to the Liouville equation for the conformal factor $\varphi$, \eqref{Liouville}, together with algebraic equations for the other fields. When the Riemann surface contains punctures or conical singularities one has to add local source terms to the right hand side of \eqref{Liouville} as in \eqref{Liouvillesource}. Given a solution to the Liouville equation, all other fields are fixed in terms of $\varphi$. The field strengths are given by \eqref{eq:F1F2M5}, the scalars $\lambda_1$ and $\lambda_2$ are as in \eqref{eq:lam12M5sol}, and the constants appearing in the metric are given by \eqref{eq:ef0eg0M5}. As discussed in Section~\ref{sec:puncbranes}, the parameters $a_1$ and $a_2$ consist of a global part and a local part accounting for the local contribution of the punctures, see \eqref{eq:aglobloc}. We can identify the global geometric contributions $a^1_{\rm global}$ and $a^2_{\rm global}$ with the Chern numbers of the line bundles in the smooth case and thus $a^1_{\rm global}+a^2_{\rm global} = 2\mathbf{g}-2$. The parameter $\mathfrak{z}$ defined in Section~\ref{sec:M5branes} is related to the parameter $z$ as
\begin{equation}
	z = \f{(2g-2)\mathfrak{z}+(a^1_{\rm local})-(a^2_\text{\rm local})}{a_1+a_2}\,.
\end{equation}
We now explicitly uplift such a solution around a puncture and provide an interpretation of the local contribution of each puncture.

\subsubsection{$\mathcal{N}=2$ and Gaiotto-Maldacena} 
\label{subsec:GM}

Locally, around a single puncture we can without loss of generality restrict ourselves to the situation $z=1$, i.e. by performing a gauge transformation we can always locally put $a_2=0$. In this case supersymmetry is enhanced to $\mathcal{N}=2$ and the solution simplifies considerably. The various fields are given by
\begin{equation}\label{puncsol}
\lambda_2 = -\f23\lambda_1 = \f15\log 2\,, \qquad F_{x_1x_2}^{(1)} = \frac{1}{8g} \e^{\varphi}\,,\qquad F_{x_1x_2}^{(2)} = 0\,. 
\end{equation}
where $\varphi$ still solves the Liouville equation \eqref{Liouville}. We can uplift this solution using the uplift formulae summarized in Appendix~\ref{app:upliftM5} to find the metric
\begin{multline}\label{MN}
\dd s_{11}^2 = \f{1}{2g^2}\tilde{\Delta}^{1/3}\dd s_{\text{AdS}_5}^2 + \f{\tilde{\Delta}^{-2/3}}{4g^2}\Big(\tilde{\Delta}\dd s_{\Sigma}^2+\tilde{\Delta}\dd \alpha^2 + \cos^2\alpha(\dd\beta^2 + \sin^2\beta \dd\phi_2^2)\\
+ 2\sin^2\alpha(\dd\phi_1+2m A^1)^2\Big)\,,
\end{multline}
where $\tilde{\Delta}=1+\cos^2\alpha$ and $\dd s^2_{\Sigma} = \e^\varphi(\dd x_1^2+\dd x_2^2)$ is the metric on the Riemann surface with unit Gaussian curvature. When we make the following coordinate change
\begin{equation}
\cos^2\alpha \rightarrow \f{y^2}{N^2}\,,
\end{equation}
we can fit this solution in the analysis of Lin-Lunin-Maldacena \cite{Lin:2004nb} to find the following eleven-dimensional supergravity background
\begin{equation}\label{LLM}
\begin{aligned}
\dd s_{11}^2&= \left(\f{\pi \ell_p^3}{2}\right)^{2/3} \e^{2  \tilde \lambda}\left( 4
\dd s_{AdS_5}^2+y^2\e^{-6{\tilde\lambda}} \dd {\tilde\Omega}_2^2+
\dd s_4^2\right)  \\
\dd s_4^2&= \f{4}{1 - y \partial_y D} (\dd \chi+v_i\dd x^i)^2 +
\f{- \partial_y D}{y}  \left[ \dd y^2 + \e^{D}
(\dd x_1^2 + \dd x_2^2) \right]
\\
v_i &= -\f12 \epsilon_{ij} \partial_j D ~~~~~~~ v= v_i \dd x_i
\\
\e^{-6\tilde \lambda}&= - \f{\partial_y D}{y(1 -  y \partial_y D)}
\\ 
G_4 = &  \f{\pi \ell_p^3}{2} F_2 \wedge \dd \Omega_2 \\
F_2 =  &  2 \left[ ( \dd t + v) \dd ( y^3 \e^{- 6 \tilde \lambda } ) + y ( 1 - y^2 \e^{ - 6 \tilde \lambda} )
\dd  v - \f12\partial _y \e^D \dd x_1\wedge \dd x_2 \right]
\end{aligned}
\end{equation}
where the function $D$ satisfies the $\SU(\infty)$  Toda equation
\begin{equation}\label{Toda}
\left( \partial_{x_1}^2+\partial_{x_2}^2 \right)D + \partial_{y}^2\e^D =0\,,
\end{equation}
and we have used the fact that the gauge coupling constant is related to the radius of the AdS$_7$ appearing in the UV, $R_{{\rm AdS}_7} = \f{2}{g} =  (\pi N)^{1/3} \ell_p$. For our $\mathcal{N}=2$ solution \eqref{puncsol}, the function $D$ takes the form
\begin{equation}\label{7dD}
\e^D = \f{\e^\varphi}{2}\left( N^2 - y^2 \right)\,.
\end{equation}
When the Riemann surface is smooth the conformal factor $\varphi$ reduces to the constant curvature metric \eqref{concurv} and the solution \eqref{MN} reduces to the $\mathcal{N}=2$ solution of Maldacena-Nu\~nez \cite{Maldacena:2000mw}.

\subsection{Punctures in eleven dimensions}\label{subsec:punc11}

Since locally all punctures we consider preserve $\mathcal{N}=2$ supersymmetry it suffices to discuss them in the $\mathcal{N}=2$ framework of \cite{Gaiotto:2009gz}. After a discussion in eleven dimensions we will go back to seven dimensions and learn how to characterize the punctures there. We will then extend the analysis to include $\mathcal{N} = 1$ supersymmetric solutions in seven dimensions.

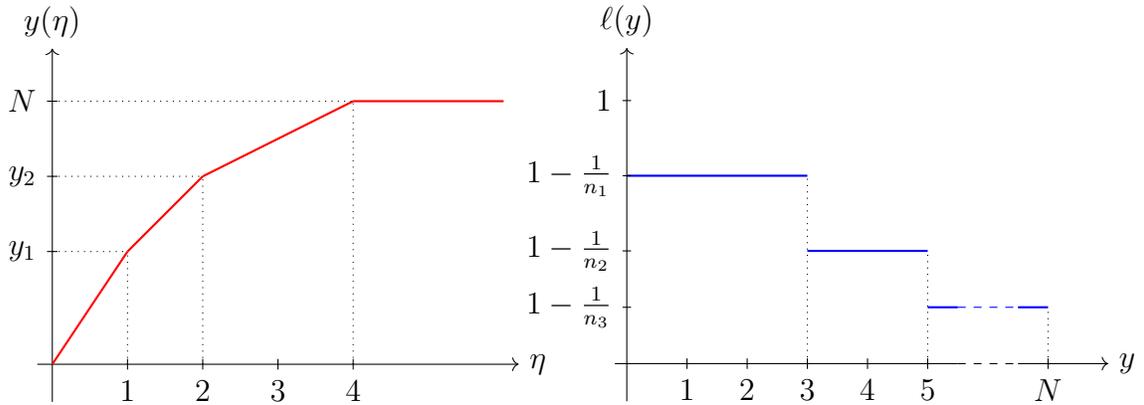
\begin{figure}[H]
	\begin{minipage}{0.45\linewidth}
		\begin{tikzpicture}
		\draw[->] (-0.2,0) -- (6.2,0) node[right] {$\eta$};
		\draw[->] (0,-0.5) -- (0,4.2) node[above] {$y(\eta)$};
		\draw[domain=0:1,thick,variable=\x,red] plot ({\x},{1.5*\x});
		\draw[domain=1:2,thick,variable=\x,red] plot ({\x},{1*\x+0.5});
		\draw[domain=2:4,thick,variable=\x,red] plot ({\x},{0.5*\x+1.5});
		\draw[domain=4:6,thick,variable=\x,red] plot ({\x},{3.5});
		\draw[smooth] (1,2pt) -- (1,-2pt) node[below] {$1$};
		\draw[smooth] (2,2pt) -- (2,-2pt) node[below] {$2$};
		\draw[smooth] (3,2pt) -- (3,-2pt) node[below] {$3$};
		\draw[smooth] (4,2pt) -- (4,-2pt) node[below] {$4$};
		\draw[dotted] (1,0) -- (1,1.5);
		\draw[dotted] (2,0) -- (2,2.5);
		\draw[dotted] (4,0) -- (4,3.5);
		\draw[dotted] (0,1.5) -- (1,1.5);
		\draw[dotted] (0,2.5) -- (2,2.5);
		\draw[dotted] (0,3.5) -- (4,3.5);
		\foreach \y/\ytext in {1.5/y_1, 2.5/y_2, 3.5/N}
		\draw[shift={(0,\y)}] (2pt,0pt) -- (-2pt,0pt) node[left] {$\ytext$};
		\end{tikzpicture}
	\end{minipage}
	\begin{minipage}{0.55\linewidth}
		\begin{tikzpicture}[xscale=0.8]
		\draw[-] (-0.2,0) -- (5.5,0);
		\draw[dashed] (5.5,0) -- (6.5,0);
		\draw[->] (6.5,0) -- (8,0) node[right] {$y$};
		\draw[->] (0,-0.5) -- (0,4.2) node[above] {$\ell(y)$};
		\draw[domain=0:3,thick,variable=\x,blue] plot ({\x},{2.5});
		\draw[domain=3:5,thick,variable=\x,blue] plot ({\x},{1.5});
		\draw[domain=5:5.5,thick,variable=\x,blue] plot ({\x},{0.75});
		\draw[domain=5.5:6.5,dashed,variable=\x,blue] plot ({\x},{0.75});
		\draw[domain=6.5:7,thick,variable=\x,blue] plot ({\x},{0.75});
		\draw[smooth] (1,2pt) -- (1,-2pt) node[below] {$1$};
		\draw[smooth] (2,2pt) -- (2,-2pt) node[below] {$2$};
		\draw[smooth] (3,2pt) -- (3,-2pt) node[below] {$3$};
		\draw[smooth] (4,2pt) -- (4,-2pt) node[below] {$4$};
		\draw[smooth] (5,2pt) -- (5,-2pt) node[below] {$5$};
		\draw[smooth] (7,2pt) -- (7,-2pt) node[below] {$N$};
		\draw[smooth] (2pt,3.5) -- (-2pt,3.5) node[left] {$1$};
		\draw[smooth] (2pt,2.5) -- (-2pt,2.5) node[left] {$1-\tfrac{1}{n_1}$};
		\draw[smooth] (2pt,1.5) -- (-2pt,1.5) node[left] {$1-\tfrac{1}{n_2}$};
		\draw[smooth] (2pt,0.75) -- (-2pt,0.75) node[left] {$1-\tfrac{1}{n_3}$};
		\draw[dotted] (3,0) -- (3,2.5);
		\draw[dotted] (5,0) -- (5,1.5);
		\draw[dotted] (7,0) -- (7,0.75);
		\end{tikzpicture}
	\end{minipage}
	\caption{piecewise linear function $y(\eta)$ and the accompanying step function $\ell(y)$.}
	\label{fig:stepfunction}
\end{figure}

Solving the Toda equation in a background with a general Riemann surface with localized punctures is hard to do in a closed form. However, locally around a puncture one can analyze the Toda equation and describe the boundary conditions for all types of regular punctures. Around an $\mathcal{N}=2$ punctures, the function $D$ satisfies the axially symmetric $\SU(\infty)$ Toda equation. By performing the transformation  \cite{Gaiotto:2009gz}
\begin{equation}\label{backlund}
	r^2 \e^D = \rho^2\,,\qquad y = \rho \partial_\rho V \equiv \dot{V}\,,\qquad \log r = \partial_\eta V \equiv V^\prime\,,
\end{equation}
we can transform the problem of solving the Toda equation to a three-dimensional axially symmetric electrostatics problem \cite{Ward:1990qt}. After this transformation the Toda equation becomes the cylindrically symmetric Laplace equation in three dimensions, 
\begin{equation}
	\ddot{V} + \rho^2 V^{\prime\prime} = 0\,.
\end{equation}
It was understood in \cite{Gaiotto:2009gz} that solutions to this equation correspond to regular geometries, up to possibly $A_{k-1}$ singularities, if and only if the line charge density $\dot{V}$ is given by a piecewise linear function with decreasing integer slopes where slope changes only occur at integer values of $\eta$ (or $y$). At points where the slope decreases by $k$ units one finds an $A_{k-1}$ singularity. Near a segment with constant slope $\dot{V}^\prime$ the potential behaves as $V\sim \dot{V}(\eta)\log \rho$. This implies that $\log r = V^\prime  = \dot{V}^\prime \log \rho$ and
\begin{equation}
	D = 2\left( \log \rho - \log r \right) = -2 \left( 1-\f{1}{\dot{V}^\prime} \right) \log r \,.
\end{equation}
This expression is valid in the range of $y$ where the slope is constant. The general boundary conditions for the Toda equation at a specific puncture thus become 
\begin{equation}\label{eq:Todabdry}
\square D = -4\pi \ell(y)\delta^{(2)}(r)
\end{equation}
where $\ell(y)$ is a piecewise constant function which only changes value at integer values of $y$. These constants decrease and take value $\left(1-\tfrac{1}{n_i}\right)$ where $n_i\in\mathbf{N}$ is the slope of the $i$th segment. This is illustrated in Figure~\ref{fig:stepfunction}.
If we have various segments with change in slope $k_i$ then the total global symmetry associated to this puncture is 
\begin{equation}
	G=\f{ \prod_i U(k_i)}{\U(1)}\,,
\end{equation}
The overall $\U(1)$ is related to the axial symmetry around the puncture which is not a real global symmetry of the system and hence is modded out.

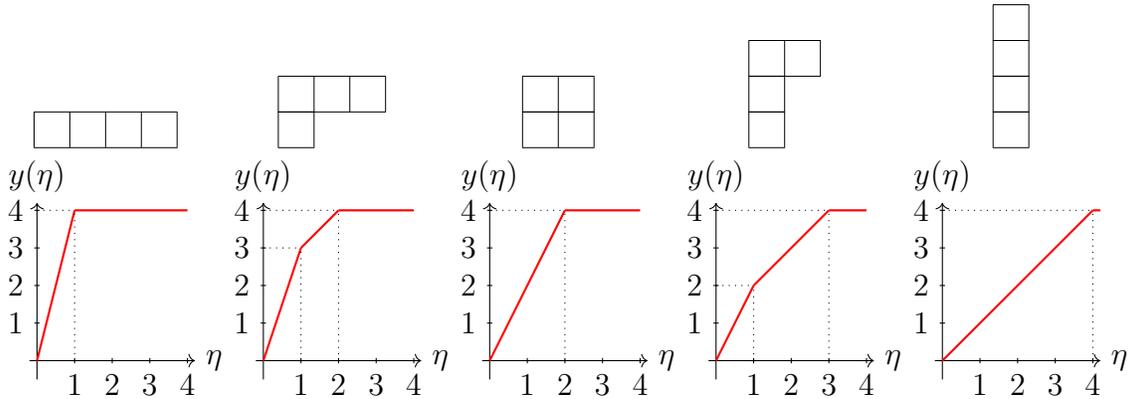
\begin{figure}[H]
	\begin{minipage}[b][][b]{0.19\linewidth}
		\centering\yng(4)\\
		\begin{tikzpicture}[scale=0.5]
		\draw[->] (-0.2,0) -- (4.2,0) node[right] {$\eta$};
		\draw[->] (0,-0.5) -- (0,4.2) node[above] {$y(\eta)$};
		\draw[domain=0:1,thick,variable=\x,red] plot ({\x},{4*\x});
		\draw[domain=1:4,thick,variable=\x,red] plot ({\x},{4});
		\draw[smooth] (1,2pt) -- (1,-2pt) node[below] {$1$};
		\draw[smooth] (2,2pt) -- (2,-2pt) node[below] {$2$};
		\draw[smooth] (3,2pt) -- (3,-2pt) node[below] {$3$};
		\draw[smooth] (4,2pt) -- (4,-2pt) node[below] {$4$};
		\draw[dotted] (1,0) -- (1,4);
		\draw[dotted] (0,4) -- (1,4);
		\foreach \y/\ytext in {1/1, 2/2, 3/3,4/4}
		\draw[shift={(0,\y)}] (2pt,0pt) -- (-2pt,0pt) node[left] {$\ytext$};
		\end{tikzpicture}
	\end{minipage}
	\begin{minipage}[b][][b]{0.19\linewidth}
		\centering\yng(3,1)\\
		\begin{tikzpicture}[scale=0.5]
		\draw[->] (-0.2,0) -- (4.2,0) node[right] {$\eta$};
		\draw[->] (0,-0.5) -- (0,4.2) node[above] {$y(\eta)$};
		\draw[domain=0:1,thick,variable=\x,red] plot ({\x},{3*\x});
		\draw[domain=1:2,thick,variable=\x,red] plot ({\x},{\x+2});
		\draw[domain=2:4,thick,variable=\x,red] plot ({\x},{4});
		\draw[smooth] (1,2pt) -- (1,-2pt) node[below] {$1$};
		\draw[smooth] (2,2pt) -- (2,-2pt) node[below] {$2$};
		\draw[smooth] (3,2pt) -- (3,-2pt) node[below] {$3$};
		\draw[smooth] (4,2pt) -- (4,-2pt) node[below] {$4$};
		\draw[dotted] (1,0) -- (1,3);
		\draw[dotted] (0,3) -- (1,3);
		\draw[dotted] (2,0) -- (2,4);
		\draw[dotted] (0,4) -- (2,4);
		\foreach \y/\ytext in {1/1, 2/2, 3/3,4/4}
		\draw[shift={(0,\y)}] (2pt,0pt) -- (-2pt,0pt) node[left] {$\ytext$};
		\end{tikzpicture}
	\end{minipage}
	\begin{minipage}[b][][b]{0.19\linewidth}
		\centering\yng(2,2)\\
		\begin{tikzpicture}[scale=0.5]
		\draw[->] (-0.2,0) -- (4.2,0) node[right] {$\eta$};
		\draw[->] (0,-0.5) -- (0,4.2) node[above] {$y(\eta)$};
		\draw[domain=0:2,thick,variable=\x,red] plot ({\x},{2*\x});
		\draw[domain=2:4,thick,variable=\x,red] plot ({\x},{4});
		\draw[smooth] (1,2pt) -- (1,-2pt) node[below] {$1$};
		\draw[smooth] (2,2pt) -- (2,-2pt) node[below] {$2$};
		\draw[smooth] (3,2pt) -- (3,-2pt) node[below] {$3$};
		\draw[smooth] (4,2pt) -- (4,-2pt) node[below] {$4$};
		\draw[dotted] (2,0) -- (2,4);
		\draw[dotted] (0,4) -- (2,4);
		\foreach \y/\ytext in {1/1, 2/2, 3/3,4/4}
		\draw[shift={(0,\y)}] (2pt,0pt) -- (-2pt,0pt) node[left] {$\ytext$};
		\end{tikzpicture}
	\end{minipage}
	\begin{minipage}[b][][b]{0.19\linewidth}
		\centering\yng(2,1,1)\\
		\begin{tikzpicture}[scale=0.5]
		\draw[->] (-0.2,0) -- (4.2,0) node[right] {$\eta$};
		\draw[->] (0,-0.5) -- (0,4.2) node[above] {$y(\eta)$};
		\draw[domain=0:1,thick,variable=\x,red] plot ({\x},{2*\x});
		\draw[domain=1:3,thick,variable=\x,red] plot ({\x},{\x+1});
		\draw[domain=3:4,thick,variable=\x,red] plot ({\x},{4});
		\draw[smooth] (1,2pt) -- (1,-2pt) node[below] {$1$};
		\draw[smooth] (2,2pt) -- (2,-2pt) node[below] {$2$};
		\draw[smooth] (3,2pt) -- (3,-2pt) node[below] {$3$};
		\draw[smooth] (4,2pt) -- (4,-2pt) node[below] {$4$};
		\draw[dotted] (1,0) -- (1,2);
		\draw[dotted] (0,2) -- (1,2);
		\draw[dotted] (3,0) -- (3,4);
		\draw[dotted] (0,4) -- (3,4);
		\foreach \y/\ytext in {1/1, 2/2, 3/3,4/4}
		\draw[shift={(0,\y)}] (2pt,0pt) -- (-2pt,0pt) node[left] {$\ytext$};
		\end{tikzpicture}
	\end{minipage}
	\begin{minipage}[b][][b]{0.19\linewidth}
		\centering\yng(1,1,1,1)\\
		\begin{tikzpicture}[scale=0.5]
		\draw[->] (-0.2,0) -- (4.2,0) node[right] {$\eta$};
		\draw[->] (0,-0.5) -- (0,4.2) node[above] {$y(\eta)$};
		\draw[domain=0:4,thick,variable=\x,red] plot ({\x},{\x});
		\draw[domain=4:4.2,thick,variable=\x,red] plot ({\x},{4});
		\draw[smooth] (1,2pt) -- (1,-2pt) node[below] {$1$};
		\draw[smooth] (2,2pt) -- (2,-2pt) node[below] {$2$};
		\draw[smooth] (3,2pt) -- (3,-2pt) node[below] {$3$};
		\draw[smooth] (4,2pt) -- (4,-2pt) node[below] {$4$};
		\draw[dotted] (4,0) -- (4,4);
		\draw[dotted] (0,4) -- (4,4);
		\foreach \y/\ytext in {1/1, 2/2, 3/3,4/4}
		\draw[shift={(0,\y)}] (2pt,0pt) -- (-2pt,0pt) node[left] {$\ytext$};
		\end{tikzpicture}
	\end{minipage}
	\caption{The Young diagrams and accompanying piecewise linear functions for $N=4$. The first diagram from the left represents a maximal puncture, the third is a $\mathbf{Z}_2$ singularity, the fourth is a minimal puncture, while the rightmost diagram corresponds to a regular point.}
	\label{fig:Younglin}
\end{figure}

Let us now relate these geometries to the punctures introduced in Section \ref{sec:M5branes}. The geometry around a puncture is fully determined by the piecewise linear function $y(\eta)$. To every such function we can associate a set of integers $n_i$ representing the slopes of the segments extending from $\eta=i-1$ to $\eta=i$. To the puncture with slopes $n_i$, we associate a Young tableau such that $n_i$ is the length of the $i$th row. We illustrate this for $N=4$ in Figure~\ref{fig:Younglin}. This provides the connection between the supergravity punctures and the correct Young tableau; indeed we see that the global symmetries preserved at each puncture matches with the discussion in Section~\ref{sec:M5branes}. 
This analysis shows that every puncture is determined by a $y$ dependent function $\eta(y)$ determining the structure of the puncture. We will not be able to capture all information contained in this function in seven dimensions but we will see that we nevertheless can extract a lot of information about the puncture solely from the seven-dimensional supergravity.

\subsection{Punctures in seven dimensions}
\label{subsec:punc7}

When we insert \eqref{7dD} in the $\SU(\infty)$ Toda equation \eqref{Toda} we obtain the Liouville equation for $\varphi$ \eqref{Liouville}. Similar to the Toda equation, finding global solutions to the Liouville equation in closed form on a general Riemann surface with punctures is hard. Nevertheless, we can learn a lot about the solutions of interest by analyzing them locally near a puncture. At a fixed value of $y$, the boundary condition of the Toda equation \eqref{eq:Todabdry} reduce to the following boundary condition for the Liouville equation
\begin{equation}\label{condef}
\varphi \sim -2(1-\tfrac{1}{n_i})\log r\,, \qquad \text{as} \quad r \rightarrow 0 \,.
\end{equation}
This is exactly the boundary condition describing a conical defect on the Riemann surface where $0<\tfrac{1}{n_i}<1$ parametrizes the opening angle at the conical singularity. We thus conclude that the different punctures in eleven dimensions correspond to conical defects on the Riemann surface of our seven-dimensional solutions, where the opening angle changes as a function of $y$. This change of opening angle goes beyond the seven-dimensional supergravity approximation and can only be treated approximately in the seven-dimensional framework. However, for some solutions -- such as $\mathbf{Z}_k$ orbifold singularities -- there is only one value of the slope in eleven dimensions and exact results can be obtained purely from seven dimensions. This kind of punctures correspond to rectangular Young tableaux and for them we can identify
\begin{equation}
	\xi = \frac{1}{n}\,.
\end{equation}
For more general punctures, where the piecewise linear function consists of more then one linear piece, we have to content ourselves with an approximate description of the eleven-dimensional puncture by specifying a single $\xi$ as the inverse of the average slope of the various segments.\footnote{For example in Figure \ref{fig:Younglin} the average slopes would be resp. $4$, $2$, $2$, $4/3$, $1$.}

Let us now go back to the supergravity solution in Section~\ref{subsec:7dM5sol} and translate our results to seven dimensions. In an $\mathcal{N}=2$ theory the field strengths $F^{(1),(2)}$ are completely fixed by the twist in terms of $\varphi$ and the contribution of the punctures manifests itself only through the volume, $\vol_{\mathbf{g},\xi_i}$, of the Riemann surface which explicitly depends on the opening angles of the conical defects,
\begin{equation}
\vol_{\mathbf{g},\xi} = \f{2\pi}{\kappa}\left(2-2\mathbf{g} - \sum_{P_i}(1-\xi_i)\right)\,.
\end{equation}
In an $\mathcal{N}=1$ theory the punctures still contribute to the volume but also the field strengths will be modified. More specifically, every puncture in an $\mathcal{N}=1$ theory is labelled by a sign $\sigma_i$ indicating in which transverse direction the puncture extends. This label indicates whether the puncture $P_i$ contributes to $a^1_{\rm local}$ or $a^2_{\rm local}$ and results in
\begin{equation}
\begin{aligned}
a^1_{\rm local} =& \sum_{\{P_i|\sigma_i = 1\}} (1-\xi_i)\,,\\
a^2_{\rm local} =& \sum_{\{P_i|\sigma_i = -1\}} (1-\xi_i)\,,
\end{aligned}
\end{equation}
where $\xi_i$ is the inverse of the average slope of the $i$th puncture and the sums above run exclusively over punctures with $\sigma_i = \pm 1$ respectively. 

\subsection{Central charges and M2 brane operators}

We can now proceed and compute the central charges $a$ and $c$ for our supergravity solutions using standard holographic results \cite{Henningson:1998gx}. The holographic conformal anomalies at leading order in $N$ are given by
\begin{equation}\label{a1}
	a = c = \frac{\pi R_{AdS_5}^3}{8G_N^{(5)}} = \left(\f{2}{g}\right)^2\frac{2\pi^3 R_{AdS_5}^3R_{S^4}^4 \vol_{\mathbf{g},\xi}\e^{\varphi_0}}{3G_N^{(11)}}\,,
\end{equation}
where $\ell_p$ is the Planck length in eleven dimensions, $G_N^{(5)}$ and $G_N^{(11)} = 16 \pi^7 \ell_p^9$ are the five- and eleven-dimensional Newton constants, and $R_{AdS_5} = \f{2\e^{f_0}}{g} = (\pi N)^{1/3} \ell_p \e^{f_0}$ and $R_{S^4} =\frac{4}{g}= 2(\pi N)^{1/3} \ell_p$ are the radii of AdS$_5$ and the four-sphere respectively. Inserting this in \eqref{a1} we obtain
\begin{equation}
	a = c = \f{2\vol_{\mathbf{g},\xi}}{3\pi}N^3 \e^{3f_0+\varphi_0}\,.
\end{equation}
Using the expressions \eqref{eq:ef0eg0M5} for $f_0$ and $\varphi_0$, we find
\begin{equation}\label{agrav}
	a = c = -\f{\kappa\,\vol_{\mathbf{g},\xi}}{2\pi} \, N^3 \f{\kappa(1-9{z}^2) + \left(1+3{z}^2\right)^{3/2}}{96{z}^2}\,.
\end{equation}
This matches exactly the large $N$ result obtained from the anomaly polynomial in Section~\ref{sec:M5branes}. Furthermore when we remove all the punctures  we recover the result for the central charges of M5-branes wrapped on a smooth Riemann surface in \cite{Bah:2012dg}. If all the punctures are minimal, the leading order result for the large $N$ central charges is not affected. This is indeed as expected since minimal punctures correspond to the addition of hypermultiplets to the dual quiver gauge theory. These indeed only contribute at order $N^2$ and will thus not be visible at leading order in $N$.

For $\mathbf{g}=0$ we find that the central charges are negative for $\sum_{i}(1-\xi_i) < 2$ indicating the presence of enhanced global symmetries rendering our $a$-maximisation calculation invalid. Indeed it is a well-known fact that one can use this leftover global symmetry to fix the positions of three labelled points on the sphere. The minimal case with no extra symmetries is the sphere with three maximal punctures which corresponds to a $\mathcal{T}_N$ building block and indeed the large $N$ anomalies \eqref{agrav} correctly describe those of a single $\mathcal{T}_N$. 

Similarly for the torus without punctures the anomaly vanishes implying there is a leftover symmetry which can be used to fix the position of a single labelled point. Indeed for a torus with one or more punctures there is no extra symmetry left and we find non-vanishing positive anomalies. For a torus with one or more punctures which contribute at order $N^3$ we again find agreement with the anomaly polynomial. When the torus has only minimal punctures, we correctly reproduce the expected $N^2$ scaling of the anomalies. However, the anomaly thus obtained does not exactly match the one obtained from the anomaly polynomial due to the presence of other 'subleading' terms which also scale as $N^2$ which are not included in our analysis. We will discuss this setup in more detail in Section~\ref{subsec:g1Laba}.

Our supergravity background describe the so-called wrapped brane geometries, see \cite{Gauntlett:2003di} for a review, where the curve $\mathcal{C}$ is a supersymmetric cycle to which one can associate a canonical BPS operator corresponding to an M2-brane wrapping this cycle \cite{Gaiotto:2009gz}. The dimension of this operator is given by the energy of the wrapped M2-brane, which we can compute at leading order in $N$ using the supergravity dual. The result is 
\begin{equation}\label{dimM2}
	\Delta(\mathcal{O}_{M2}) = \f{2\vol_{\mathbf{g},\xi_i}}{\pi} N \e^{f_0+\varphi_0-2(\lambda_1+\lambda_2)} = -\kappa \f{\vol_{\mathbf{g},\xi_i}}{4\pi}N\left( 1- \f{\kappa}{2} \sqrt{1+3{z}^2} \right)\,,
\end{equation}
which at large $N$ and for specific values of the parameters indeed matches with the known results in \cite{Gaiotto:2009gz,Benini:2009mz,Bah:2012dg}. In Section~\ref{subsec:Dualconfman} we show how to compute this dimension directly in the holographic dual SCFT.

\subsection{Marginal deformations}

From our supergravity solutions we can also compute the exactly marginal deformations of the dual SCFT. Doing this calculation rigorously requires computing the spectrum of  KK excitations around the eleven-dimensional solutions presented above and identify the scalar modes with vanishing mass. This is a hard task which we do not know how to do in general. Nevertheless, there is a natural set of massless modes in the solutions which we believe to exhaust the list and thus span the superconformal manifold in the dual SCFT. To illustrate this we focus on the case of Riemann surfaces with $n$ indistinguishable minimal punctures. One in principle can straightforwardly extend this computation to compute the dimension of the moduli space of a Riemann surface with different kinds of punctures but the computations quickly become very tedious.

The first set of marginal deformations is given by the moduli space of the algebraic curve $\mathcal{C}$ denoted by $\mathcal{M}_{\mathbf{g},n}$. For $\mathbf{g}>1$ these enter the construction in the action of the Fuchsian group $\Gamma$ on the hyperbolic plane $\mathbf{H}$ spanned by $(x_1,x_2)$. For $\mathbf{g}>1$ the dimension of this moduli space is $\dim_{\mathbf{C}} \mathcal{M}_{\mathbf{g},n} = 3\mathbf{g}-3+n$.\footnote{For the two-sphere $\dim_{\mathbf{C}}\mathcal{M}_{0,0}=\dim_{\mathbf{C}}\mathcal{M}_{0,1}=\dim_{\mathbf{C}}\mathcal{M}_{0,2}=0$. For $n\geq3$ punctures $\mathcal{M}_{0,n} = n-3$. A torus with $n$ punctures has $\dim_{\mathbf{C}}\mathcal{M}_{1,n}=n+1$.} These moduli correspond to the complex structure deformations of the curve and can be identified with the complex gauge couplings in the dual field theory. For $\mathcal{N}=2$ solutions these are the only moduli present compatible with supersymmetry.

For $\mathcal{N}=1$ solutions a second set of exactly marginal deformations arises from the freedom to shift the gauge fields by a flat connection on $\mathcal{C}$, 
\begin{equation}
	A^{(1)} \rightarrow A^{(1)} + A_\text{flat}\,,\qquad\qquad A^{(2)} \rightarrow A^{(2)} - A_\text{flat}\,.
\end{equation}
Such a shift leaves the BPS equations invariant. For generic $\mathbf{g}$ and $n$ there are $2\mathbf{g}+n$ flat $\U(1)$ connections. Additionally every puncture contributes an additional real parameter corresponding to the $\mathbf{CP}^1$ worth of direction inside the two-dimensional fiber over $\mathcal{C}$, see \cite{Benini:2009mz} for a discussion on this extra modulus. Therefore the complex dimension of the supersymmetric conformal manifold of the dual SCFT is
\begin{equation}\label{dimMc}
\dim_{\mathbf{C}}\mathcal{M}_C = 4\mathbf{g}-3+2n\,.
\end{equation}
In the special case when $z=0$ there is an overall $\SU(2)_\mathcal{F}$ flavor symmetry leading to additional marginal deformations related to $\SU(2)_\mathcal{F}$ Wilson lines on $\mathcal{C}$, see \cite{Benini:2009mz}. There are $6\mathbf{g}-6+3n$ such flat connections and thus the dimension of the conformal manifold becomes $\dim_{\mathbf{C}}\mathcal{M}_C = 6\mathbf{g}-6+3n$.

\section{Dual quiver gauge theories} 
\label{sec:quivers}

In this section we describe the four-dimensional quiver gauge theories dual to the $\mathcal{N}=1$ AdS$_5$ solutions studied in the previous sections. As in the gravitational case we restrict ourselves to $\SU(N)$ quiver gauge theories originating from the $A_{N-1}$ type $\mathcal{N}=(2,0)$ theory. It is possible to study other gauge groups and quivers originating from $D_N$ or even $E_n$ type theories along similar lines by using the results in \cite{Gaiotto:2009we,Tachikawa:2009rb,Chacaltana:2010ks,Chacaltana:2011ze,Chacaltana:2014jba,Chacaltana:2017boe,Chacaltana:2018zag}. We also restrict our analysis to minimal, or simple, punctures and compute the 't Hooft anomalies, the dimensions of protected operators and the dimension of the conformal manifold for these quiver gauge theories. Minimal punctures correspond in the quiver gauge theory to hypermultiplets connecting adjacent gauge groups. This is the simplest case to analyze but one can also study more general punctures by adding more intricate quiver tails \cite{Gaiotto:2009we,Chacaltana:2010ks,Agarwal:2014rua}. We emphasize that many of the results in this section have appeared in the literature before or can be derived in a straightforward manner following the discussion in \cite{Benini:2009mz,Bah:2011je,Bah:2011vv,Bah:2012dg,Gadde:2013fma,Bah:2013aha,Agarwal:2014rua,Bah:2017gph}.\footnote{To the best of our knowledge, the relation between the $L^{aba}$ SCFTs and theories of class $\mathcal{S}$ described in Section~\ref{subsec:g1Laba} has not appeared explicitly in the literature before.} Nevertheless we believe that the summary below serves a useful purpose to illustrate the salient features of our construction. 

\subsection{Setup and symmetries}

\begin{figure}[H]
	\centering
	\includegraphics[scale=0.5]{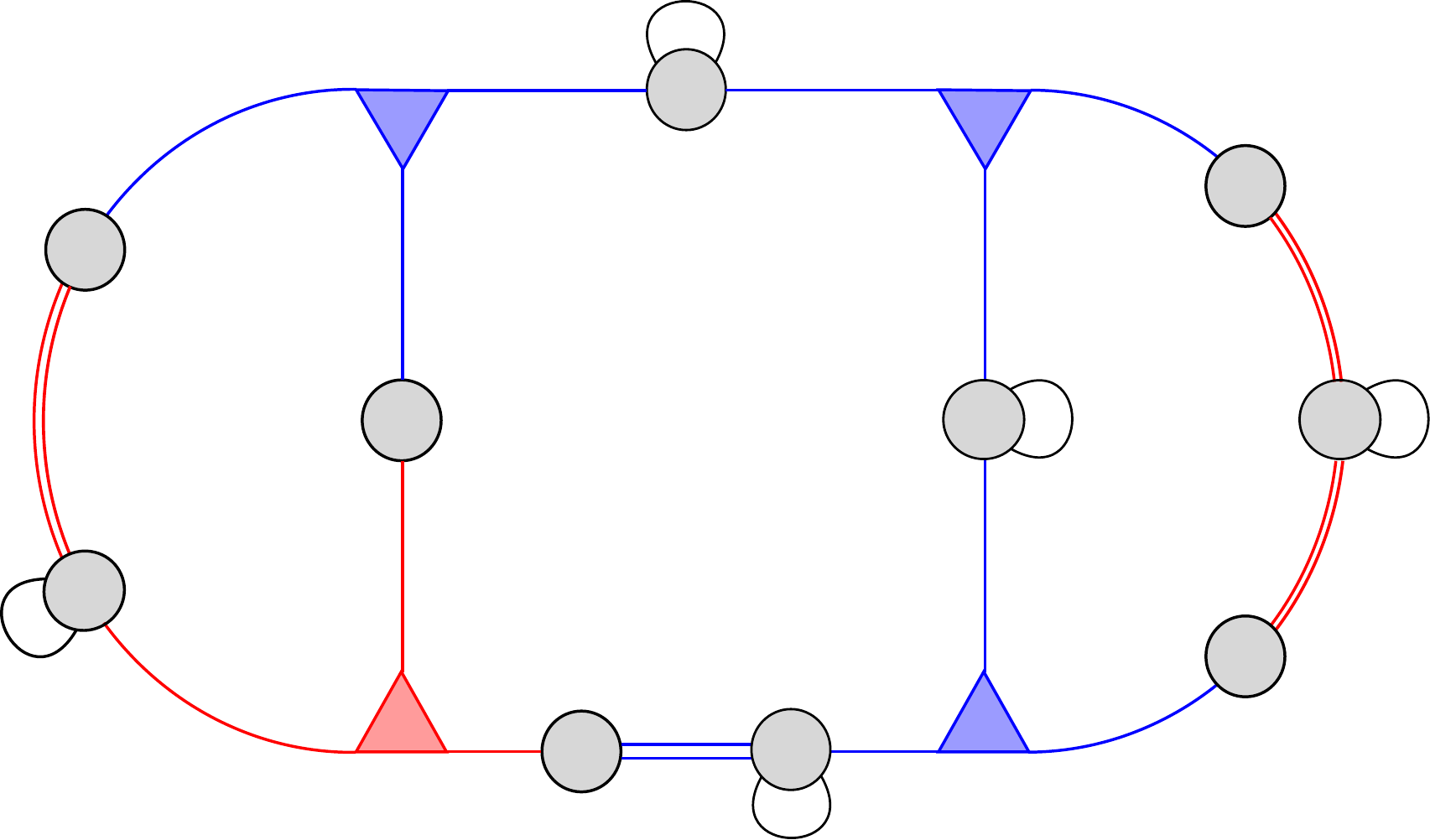}
	\caption{An example of a genus $\mathbf{g}=3$ quiver.}
	\label{fig:GeneralQuiver}
\end{figure}

The quiver gauge theories we consider are constructed from $\mathcal{T}_N$ building blocks connected by strands of linear quivers build from vector and hyper multiplets. This type of quiver is sketched in Figure \ref{fig:GeneralQuiver} which should be interpreted as follows:
\begin{itemize}
	\item Nodes without ears denote $\mathcal{N}=1$ vector multiplets with $\SU(N)$ gauge groups. 
	\item Nodes with ears denote $\mathcal{N}=2$ vector multiplets with $\SU(N)$ gauge groups. The ear represents the adjoint chiral.
	\item Double lines between two vector multiplets denote hypermultiplets in the bifundamental representation of the two adjacent gauge groups.
	\item $\mathcal{T}_N$ building blocks are denoted by triangles which are connected to three vector multiplets.
	\item The coloring of the matter field is associated to a $\mathbf{Z}_2$ valued label $\sigma_i$. Matter field with $\sigma_i = 1$ are colored red and those with $\sigma=-1$ are blue. 
\end{itemize}
A general quiver of genus $\mathbf{g}$ contains $2\mathbf{g}-2$ $\mathcal{T}_N$ building blocks combined with $3\mathbf{g}-3$ strands built out of vector and hypermultiplets. Every such linear quiver is built out of $n_i$ hypermultiplets and $n_i+1$ vector multiplets. Consequently, the total number of hypermultiplets is $n=\sum_{i}n_i$ and the number of vector multiplets is $v= 3\mathbf{g}-3+n$. Let us denote the number of $\mathcal{N}=2$ vectors by $v_2$, analogously $v_1$ is the number of $\mathcal{N}=1$ vectors.

The $\mathcal{T}_N$ theory was proposed in \cite{Gaiotto:2009we} as the low-energy theory coming from $N$ M5-branes wrapping a trice punctured sphere, see \cite{Tachikawa:2015bga} for a review. It is an $\mathcal{N}=2$ building block with $\SU(N)^3\times\SU(2)_R\times\U(1)_r$ global symmetry. Except for the case where $N=2$\footnote{In this case the $\mathcal{T}_2$ theory reduces to eight free chiral multiplets transforming in the trifundamental of the $\SU(2)^3$ global symmetry.} there is no known weakly coupled Lagrangian description for these theories. The spectrum of the $\mathcal{T}_N$ theory includes Higgs branch operators $\mu_a$ with $a=1,2,3$ called moment maps; one triplet for each $\SU(N)$ flavor group. Each such operator has dimension two and transforms in the adjoint of one of the $\SU(N)$ factors. Additionally there are also Coulomb branch operators $u_k^{(i)}$ with $k=3,\dots, N$ and $i=1,\dots,k-2$ associated to each $\mathcal{T}_N$ with dimension $k$. Finally there are dimension $N-1$ operators $Q$ and $\widetilde{Q}$ transforming, respectively, in the $(\mathbf{N},\mathbf{N},\mathbf{N})$ and $(\overline{\mathbf{N}},\overline{\mathbf{N}},\overline{\mathbf{N}})$ representation of $\SU(N)^3$. To each hypermultiplet we can also associate a triplet of moment map operators $\mu$ transforming in the adjoint of the $\SU(N)$ flavour symmetry. 

Since in general we consider $\mathcal{N}=1$ quiver gauge theories, it is convenient to think of $\mathcal{N}=2$ vector multiplets, hypermultiplets and $\mathcal{T}_N$ theories as $\mathcal{N}=1$ building blocks. An $\mathcal{N}=2$ vector multiplet can be thought of as a $\mathcal{N}=1$ vector with an additional adjoint chiral. A $\mathcal{T}_N$ on the other hand can be though of as a $\mathcal{N}=1$ building block with an additional $\U(1)$ flavour symmetry. Finally every hypermultiplet consist of two chiral multiplets in conjugate representations $H_i = \{ q_i,\tilde{q}_i \}$. In terms of these constituent fields the moment map triplet of the hypermultiplet takes the form $\mu_i^+ = q_i \tilde{q}_i$, $\mu_i^0=|q_i|^2-|\tilde{q}_i|^2$ and $\mu^-_i=(\mu_i^+)^*$. When expressing the theory as a $\mathcal{N}=1$ theory the generators of the $\mathcal{N}=2$ R-symmetry decompose into a generator for the $\mathcal{N}=1$ superconformal R-symmetry and a generator for an extra $\U(1)$ flavour symmetry,\footnote{See Appendix~\ref{app:SCFTtrivia} for more detail on our SCFT conventions.}
\begin{equation}
R_{\mathcal{N}=1}=\f{1}{3}R_{\mathcal{N}=2}+\f43 I_3\,, \quad J=R_{\mathcal{N}=2}-2I_3\,.
\end{equation}
where $I_3$ is the Cartan of $\SU(2)_R$. When the supersymmetry is broken to $\mathcal{N}=1$, $R_{\mathcal{N}=1}$ will no longer be in the same multiplet as the stress tensor and the superconformal R-symmetry at the IR fixed point may potentially mix with $\U(1)$ flavor symmetries. 

Now that we have introduced all the building blocks we can use them to construct generalized quiver gauge theories by gauging the various $\SU(N)$ global symmetries. Every $\mathcal{N}=2$ gauging introduces a superpotential term of the form
\begin{equation}
\mathcal{W}_{\mathcal{N}=2}^{\text{gauging}} = \Tr\phi_i(\mu_i+\mu_{i+1})\,,
\end{equation}
where the $\mu_{i,i+1}$ are the moment maps belonging to the adjacent matter building blocks. This superpotential breaks all the baryonic symmetries which are otherwise present. A general quiver (like the one in Figure \ref{fig:GeneralQuiver}) has $\mathcal{N}=1$ supersymmetry. However, in addition to $\mathcal{N}=1$ supersymmetry such quivers possess large amounts of global symmetries. We always have an overall $\U(1)_R$ R-symmetry and additionally for each hyper, $\mathcal{T}_N$ and adjoint chiral there is an associated $\U(1)$. We denote the $\U(1)$ symmetries acting on the hypers and $\mathcal{T}_N$ blocks by $J_i$ and the ones acting on the adjoint chiral by $F_i$. The full global symmetry is thus $\U(1)^{v_2+2\mathbf{g}-2+n}\times \U(1)_R$. However some of these symmetries are anomalous. Each gauge group, except for one global combination, provides one anomaly constraint so we end up with $2\mathbf{g}-2+v_2$ anomaly-free $\U(1)$ global symmetries.\footnote{This result is only valid for quivers with $\mathbf{g}>1$, for $\mathbf{g}=1$ we find $v_2+1$ anomaly-free $\U(1)$s.}

We can consider such gauge theories without extra superpotential terms. However, we expect these quivers to break up in smaller quivers in the IR \cite{Bah:2013aha}. Indeed, the one-loop beta-functions for the gauge group couplings are 
\begin{equation}
b_0(V_i^{\mathcal{N}=2}) = 0\,, \qquad b_0(V_i^{\mathcal{N}=1}) = - N \,.
\end{equation}
The gauge couplings for the $\mathcal{N}=2$ gauge groups are marginal and without superpotential, they should be marginally irrelevant \cite{Green:2010da}. As a result, the $\mathcal{N}=2$ gauge groups are non-dynamical in the IR and the quiver will break up at these sites. 

We are more interested in finding situations where the IR dynamics of the gauge theory is non-trivial. We expect that for an appropriate choice of the superpotential the IR physics is governed by an $\mathcal{N}=1$ SCFT. By adding specific superpotential terms we can prevent  the quivers from breaking apart in the IR. At $\mathcal{N}=1$ sites we turn on 
\begin{equation}\label{N1W}
\mathcal{W}_i^{\mathcal{N}=1} = \alpha_i \Tr \mu_{i-1}\mu_i\,,
\end{equation}
where $\alpha_i$ are arbitrary complex numbers. At $\mathcal{N}=2$ sites on the other hand we turn on the superpotential
\begin{equation}\label{N2W}
\mathcal{W}_i^{\mathcal{N}=2} = \beta^L_{i} \Tr \phi_i\mu_{i-1}+\beta^R_{i} \Tr \phi_i\mu_i \,,
\end{equation} 
where $\beta^{L,R}_{i}$ are complex numbers. The superpotentials \eqref{N1W} and \eqref{N2W} always preserve the $\U(1)_R$ R-symmetry $R_0$ generated by
\begin{equation}\label{eq:R0def}
	R_0 = R_{\mathcal{N}=1}+ \f16 \sum_{i} J_i\,.
\end{equation} 
To see if these superpotentials leave another anomaly-free $\U(1)$ unbroken we first need to understand how the chiral anomaly is cancelled locally at every site. At the $i$th node of the quiver the combination $J_{i-1}-J_i$ is always anomaly-free. When this site also contains an adjoint chiral there is a second anomaly-free $\U(1)$ with generator $J_{i-1}+ J_i - 2F_i$. Note that $\Tr(J_{i-1}-J_i)=0$ so this global symmetry is baryonic and will by itself be broken at $\mathcal{N}=2$ sites by the superpotential \eqref{N2W}. However, by combining it with the $\U(1)$'s coming from the adjoint chirals we are able to construct a non-baryonic anomaly-free $\U(1)$ global symmetry.

In order to construct such a global $\U(1)$ we can assign to all matter multiplets, hypermultiplets and $\mathcal{T}_N$ blocks, a sign $\sigma_i$ as indicated by the colouring in Figure~\ref{fig:GeneralQuiver}. When crossing an $\mathcal{N}=1$ vector the sign of two neighbouring matter fields flips. When crossing an $\mathcal{N}=2$ vector multiplet the sign remains unchanged. Not every quiver configuration allows for such an assignment of $\mathbf{Z}_2$ label. If a quiver does not allow the assignment we expect that it will  flow to the universal $\mathcal{N}=1$ IR fixed point discussed in \cite{Tachikawa:2009tt}. If a quiver does allow for a consistent sign assignment complying to these rules we have an additional $\U(1)$ flavour symmetry with generator
\begin{equation}\label{eq:U1F}
\mathcal{F} = \f12\sum_{i=1}^n \left( \sigma_i J_i - (\sigma_{i-1}+\sigma_i)F_i \right)\,.
\end{equation}
This is the only anomaly-free flavour $\U(1)$ preserved by the superpotential terms \eqref{N1W} and \eqref{N2W}. This setup and the rules for finding the global flavor $\U(1)$ follow the construction in \cite{Bah:2011je,Bah:2012dg,Bah:2013aha}.

One might additionally want to add the superpotential term 
\begin{equation}\label{schuppenzot}
\mathcal{W}^{\spadesuit}_i = \gamma_i (\mu_i)^2 \,,
\end{equation}
where the $\gamma_i$ are arbitrary complex numbers. However, when one of the $\gamma_i \neq 0$ the superpotential \eqref{schuppenzot} breaks the extra flavour symmetry $\mathcal{F}$. As we will discuss in the next section, when this happens the theories always flow to the same universal IR SCFT as in \cite{Tachikawa:2009tt}. In the following we tune all $\gamma_i$ to zero and consider quivers gauge theories with the extra $\U(1)$ global symmetry in \eqref{eq:U1F}. These theories allow for interesting new IR SCFTs since the anomaly-free flavor symmetry can mix with the R-symmetry in \eqref{eq:R0def}. To determine the correct $\mathcal{N}=1$ superconformal $R$-symmetry one then has to take the linear combination
\begin{equation}\label{Rtrial}
R_\epsilon = R_0 + \epsilon \mathcal{F}\,,
\end{equation} 
and fix the real number $\epsilon$ using $a$-maximisation \cite{Intriligator:2003jj}.

\subsection{IR dynamics} \label{subsec:IRdynamics}

Now that we have set the stage let us study the IR dynamics of the $\mathcal{N}=1$ quiver gauge theories introduced above. the upshot is that whenever the quiver allows for a consistent assignment of the labels $\sigma_i$, we find interacting SCFTs dual to the gravity solutions described in Section~\ref{sec:punctures}. Since the quivers with $\mathbf{g}=1$ do not contain any $\mathcal{T}_N$ building blocks we first focus on the generic situation with $\mathbf{g}>1$ and discuss $\mathbf{g}=1$ in Section~\ref{subsec:g1Laba}.  For a quiver of genus $\mathbf{g}>1$ we have $2\mathbf{g}-2$ $\mathcal{T}_N$ building blocks and $3\mathbf{g}-3$ strands of linear quiver. We have $p$ $\mathcal{T}_N$ blocks with positive sign and $q$ with negative sign. Similarly, we have $x$ hypers with positive sign and $y$ with negative sign. If the theory flows to an IR SCFT we can determine the IR superconformal R-symmetry using $a$-maximisation \cite{Intriligator:2003jj}. We can compute the $a$ and $c$ anomaly and determine the dimensions of the chiral operators. The central charges $a$ and $c$ are given by the 't Hooft anomalies associated with the superconformal $R$-symmetry $R_{\mathcal{N}=1}$ as in \eqref{acRanom}. 

Our quivers admit a one-parameter family of R-symmetries which are linear combinations of the UV R-symmetry $R_0$ and the global flavour symmetry $\mathcal{F}$ as in \eqref{Rtrial}. For each $R_\epsilon$ we can compute the trial central charge $a(\epsilon)$. The superconformal R-symmetry maximizes this function $a(\epsilon)$ and in this way uniquely determines the value of $\epsilon$. The charges of the superfields under $R_\epsilon$ are
\begin{equation}\label{charges}
R_\epsilon(q_i)=R_\epsilon(\tilde{q}_i)=\f12 (1+\epsilon \sigma_i)\,,\quad \text{and}\quad R_\epsilon(\phi_i)=1-\f12 \epsilon (\sigma_{i-1}+\sigma_i)\,.
\end{equation}
Consequently, the 't Hooft anomalies for the $i$th hypermultiplet are
\begin{equation}
\tr R_\epsilon(H_i) = N^2 (\epsilon\sigma_i-1)\,, \qquad \tr R_\epsilon(H_i)^3 = \f14 N^2 (\epsilon\sigma_i-1)^3\,,
\end{equation}
while for the $i$th vector multiplet we find
\begin{equation}
\begin{aligned}
\tr R_\epsilon(V_i) =& (N^2-1)\left[ 1-\f12\epsilon(\sigma_{i-1}+\sigma_i) \right]\,,\\
\tr R_\epsilon(V_i)^3 =& (N^2-1)\left[ 1-\f18 \epsilon^3 (\sigma_{i-1}+\sigma_i)^3 \right]\,.
\end{aligned}
\end{equation}
For a single $\mathcal{T}_N$ we have
\begin{equation}
\begin{aligned}
\tr R_{\epsilon}(\mathcal{T}_{Ni})=&\frac{1-\sigma_i\epsilon}{2}\tr R_{\mathcal{N}=2}^3\,,\\
\tr R_{\epsilon}(\mathcal{T}_{Ni})^3 =& \f{(1-\sigma_i\epsilon)^3}{8}\tr R_{\mathcal{N}=2}^3 + \f32 (1-\sigma_i\epsilon)(1+\sigma_i\epsilon)^2\Tr R_{\mathcal{N}=2}I_3^2\,,
\end{aligned}
\end{equation}
where
\begin{equation}
\begin{aligned}
\tr R_{\mathcal{N}=2}I_3^2 =& \f{1}{12}(6-N-9N^2+4N^3)\\
\tr R^3_{\mathcal{N}=2} =& 2+N-3N^2\,.
\end{aligned} 
\end{equation}
In the equations above the trace $\tr$ denotes the sum over all chiral fermions in the $i$th hyper, vector or $\mathcal{T}_N$, together with the trace over the gauge indices. 

We can now write the anomaly of one strand by summing over all fields in the linear piece to obtain\footnote{Here we sum over a linear quiver with $l_i+1$ vectors, $l_i$ hypers and 2 $\mathcal{T}_N$ blocks at the endpoints. The $\mathcal{T}_N$ contribution is considered separately, only the sign coming from the $\mathcal{T}_N$ is used.}
\begin{equation}
\begin{aligned}
\Tr R_\epsilon(H) =& N^2 l_i (\eta_i\epsilon-1)\,,\\
\Tr R_\epsilon(H)^3 =& \f{N^2 l_i}{4}\left( \eta_i\epsilon\left(3+\epsilon^2\right) - \left(1+3\epsilon^2\right) \right)\,,
\end{aligned}
\end{equation}
for the hypermultiplets and 
\begin{equation}
\begin{aligned}
\Tr R(\epsilon)(V) =& \left( N^2-1 \right)\left[ l_i+1-\epsilon(l_i\eta_i+\kappa_i) \right]\,,\\
\Tr R(\epsilon)(V)^3 =& \left( N^2-1 \right)\left[ l_i+1-\epsilon^3( l_i \eta_i+\kappa_i) \right]\,,
\end{aligned}
\end{equation}
for the vector multiplets. Here the trace $\Tr$ denotes the sum over all chiral fermions of the $i$th linear strand of the quiver together with the trace over gauge indices. In the expressions above we have introduced the new parameters
\begin{equation}
\eta_i = \f{1}{l_i}\sum_{i=1}^{l_i} \sigma_i= \f{x_i-y_i}{l_i}\,, \qquad \qquad \text{and}\qquad \qquad \kappa_i=\f{\sigma_a+\sigma_b}{2}\,,
\end{equation} 
where $\sigma_a$ and $\sigma_b$ are the signs of the $\mathcal{T}_N$ block adjacent to the linear strand. Now summing over all linear strands and $\mathcal{T}_N$ blocks and inserting the result in \eqref{acRanom} results in the trial central charge
\begin{equation}
\begin{aligned}
a(\epsilon) =& \f{3}{32}\left( p A(\epsilon,N) + q A(-\epsilon,N) \right) + \frac{3}{32}\f{n N^2}{4}\left[ \eta \epsilon(3\epsilon^2+5) - (9\epsilon^2-1) \right]\\
&+\f{3}{32}(N^2-1)\left[ 2n + 3(p+q) + n\eta(\epsilon-3\epsilon^3) + \f{3}{2}(p-q)(\epsilon-3\epsilon^3) \right]\,,
\end{aligned}
\end{equation}
where we have introduced the function
\begin{equation}
A(t,N) = \left[ \f{3}{8}(1-t)^3 -\f12 (1-t) \right]\tr R^3_{\mathcal{N}=2} + \f92(1-t)(1+t)^2\tr R_{\mathcal{N}=2}I_3^2 \,,
\end{equation}
and the new parameters
\begin{equation}
n = \sum_{i=1}^{3g-3}l_i \qquad\qquad\text{and }\qquad\qquad \eta=\f{\sum_{i=1}^{3\mathbf{g}-3}l_i \eta_i}{\sum_{i=1}^{3\mathbf{g}-3}l_i}= \f{x-y}{n}\,.
\end{equation}
We also used the fact that $p+q=2\mathbf{g}-2$ and $\sum_i\kappa_i= \f{3}{2}(p-q)$. Rewriting this using the parameter $\mathfrak{z}$ defined in \eqref{eq:frakzdef} we exactly recover the result obtained from the anomaly polynomial calculation presented in Section~\ref{sec:M5branes}. Note that this agreement holds even before $a$-maximization providing strong evidence that we have identified the correct M5-brane constructions corresponding to these four-dimensional quiver theories. When all $l_i=0$ our results reduce to the ones found in \cite{Bah:2012dg} for quivers corresponding to smooth Riemann surfaces. We can now maximize the function $a(\epsilon)$ to find the IR R-symmetry and the actual central charges. In general this is a rather complicated and non-illuminating function of $z$, $\eta$, $n$ and $N$ which we refrain from presenting here. However, in the large $N$ limit, $\epsilon_{\text{max}}$ reduces to
\begin{equation}
	\epsilon_{\text{max}} = \f{1-\sqrt{1+3\mathfrak{z}^2}}{3\mathfrak{z}}\,,
\end{equation}
and we find the large $N$ central charges
\begin{equation}
a = c = (1-\mathbf{g})N^3 \f{1-9\mathfrak{z}^2-(1+3\mathfrak{z}^3)^{3/2}}{48\mathfrak{z}^2}\,.
\end{equation}
This agrees with the result from the dual gravity solution presented in \eqref{agrav}. To show this note that in the case of minimal punctures we have $\xi=\frac{N-1}{N}$ and at large $N$ the supergravity parameter $z$ in \eqref{agrav} reduces to $\mathfrak{z}$,
\begin{equation}
z=\frac{p+x(1-\xi)-q-y(1-\xi)}{2\mathbf{g}-2+n(1-\xi)} \longrightarrow \f{p-q}{2\mathbf{g}-2} = \mathfrak{z}\,.
\end{equation}
The leading order large $N$ result remains unchanged by adding minimal punctures. From the gauge theory point of view it is easy to see that this should indeed be the case since hypermultiplets only contribute at order $N^2$.

\subsection{$\mathbf{g}=1$ and $L^{aba}$}
\label{subsec:g1Laba}

\begin{figure}[H]
	\centering
	\includegraphics[scale=0.5]{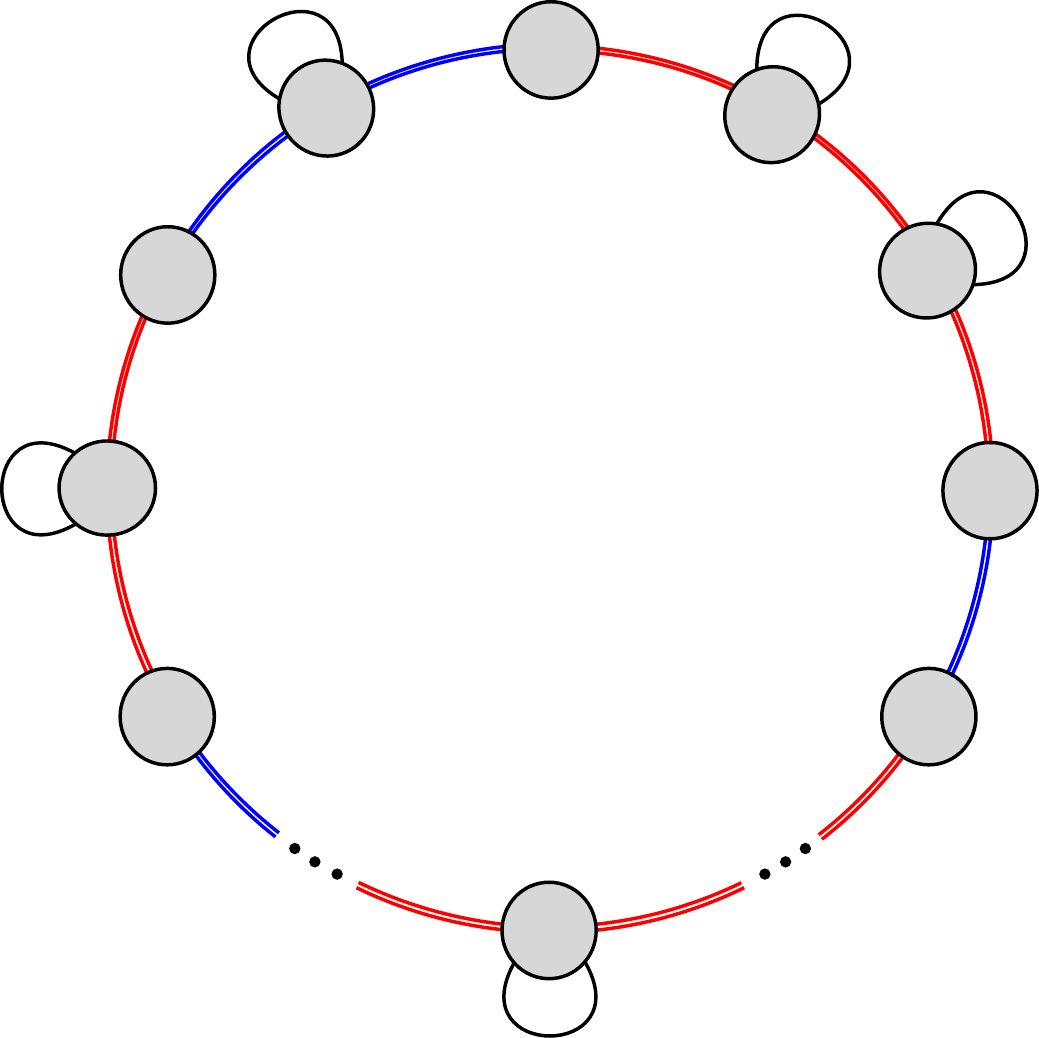}
	\caption{A Necklace quiver composed of $\mathcal{N}=1$ and $\mathcal{N}=2$ vector multiplets connected by hypermultiplets.}
	\label{fig:NecklaceQuiver}
\end{figure}

The quiver gauge theory obtained by putting the $\mathcal{N}=(2,0)$ theory on a torus with $n$ minimal punctures, without extra flux is given by a necklace quiver composed of $n$ vector and hypermultiplets and no $\mathcal{T}_N$ blocks, see Figure~\ref{fig:NecklaceQuiver}. 

We can repeat the analysis for $\mathbf{g}>1$ and obtain the 't Hooft anomalies by combining the $R$-charges of all chiral fermions in the theory. Summing over all hyper and vector multiplets in the quiver we find
\begin{equation}\label{hypers}
\begin{aligned}
\Tr R_\epsilon(H) =& N^2 n (\eta\epsilon-1)\,,\\
\Tr R_\epsilon(H)^3 =& \f{N^2 n}{4}\left( \eta\epsilon\left(3+\epsilon^2\right) - \left(1+3\epsilon^2\right) \right)\,,
\end{aligned}
\end{equation}
for the hypermultiplets and
\begin{equation}\label{vectors}
\begin{aligned}
\Tr R(\epsilon)(V) =& \left( N^2-1 \right)n(1-\epsilon \eta)\,,\\
\Tr R(\epsilon)(V)^3 =& \left( N^2-1 \right) n (1-\epsilon^3 \eta)\,,
\end{aligned}
\end{equation}
for the vector multiplets. Here the trace denotes the sum over all chiral fermions in all the hypers or vectors, of the full quiver together with the trace over gauge indices. With \eqref{hypers} and \eqref{vectors} at hand we can compute the trial central charge
\begin{equation}\label{anoma}
a(\epsilon) = \f{3n}{32}\left( \f94 N^2(1-\eta\epsilon)(1-\epsilon^2)   -2+\eta\epsilon(1-3\epsilon^2)\right)\,.
\end{equation}
This function is maximized at $\epsilon = \epsilon_{\rm max}$ where
\begin{equation}
\epsilon_{\rm max} = \f{3N^2-\sqrt{9N^4+\left(16-48N^2+27N^4\right)\eta^2}}{3\left(4-3N^2\right)\eta}\,.
\end{equation}
Inserting $\epsilon_{\rm max}$ in \eqref{anoma} we exactly reproduce the $a$ anomaly obtained in Section \ref{sec:M5branes} from the anomaly polynomial. Similarly we can exactly reproduce the $c$ anomaly by inserting $\epsilon_{\rm max}$ in the corresponding formula for $c$
\begin{equation}
c = \f{n}{32}\left( \f{27}{4} N^2(1-\eta\epsilon_{\rm max})(1-\epsilon_{\rm max}^2)   -4+\eta\epsilon_{\rm max}(5-9\epsilon_{\rm max}^2)\right)\,.
\end{equation}

In \cite{Benvenuti:2005ja,Franco:2005sm,Butti:2005sw} a large class of four-dimensional quiver gauge theories, known as $L^{abc}$, arising from D3-branes probing toric Calabi-Yau singularities were obtained. The quiver gauge theory in Figure~\ref{fig:NecklaceQuiver} corresponds to the theories $L^{aba}$ studied in Section 3.1 and 6.2.3 of \cite{Franco:2005sm}. The map between the parameters used in our work and the ones in \cite{Franco:2005sm} is 
\begin{equation}
n = a+b\,, \qquad \xi=\f{b-a}{a+b}\,.
\end{equation}
A consistency check of this identification is provided by the agreement between the large $N$ limit of the $a$-anomaly in \eqref{anoma} and the expression in Equation $(6.11)$ of \cite{Franco:2005sm}. This analysis shows that the class of $L^{aba}$ quiver gauge theories can be obtained by wrapping M5-branes on a punctured torus and therefore these theories belong to the landscape of $\mathcal{N}=1$ theories of class $\mathcal{S}$. 

\subsection{Consistency, duality, and the conformal manifold}
\label{subsec:Dualconfman}

Given the IR superconformal R-symmetry we can compute the scaling dimension of the various chiral primaries in our theory. These are given by 
\begin{equation}
\begin{array}{ll}
R_{\mathcal{N}=1}(\mu_i) = 1+ \sigma_i\epsilon_{\text{max}}\,, & R_{\mathcal{N}=1}(\phi_i^2) = 2-(\sigma_{i-1}+\sigma_i)\epsilon_{\text{max}}\,,\\
R_{\mathcal{N}=1}(u_k) = 1-\sigma_i\epsilon_{\text{max}}\,, &
R_{\mathcal{N}=1}(Q_i) = \f12(N-1)(1+\sigma_i\epsilon_{\text{max}})\,.
\end{array}
\end{equation}
The maximizing value $\epsilon_{\text{max}}$ takes values between $-\f13\leq \epsilon_{\text{max}} \leq \f13$ where the lower and upper bound are attained for $\eta=\mathfrak{z}=1$ and $\eta=\mathfrak{z}=-1$. Using this we find that the unitarity bound
\begin{equation}
\Delta = \f32 R_{\mathcal{N}=1} \geq 1 \,,
\end{equation}
is always satisfied. Moreover, the Hofman-Maldacena bound  \cite{Hofman:2008ar} for $\mathcal{N}=1$ SCFTs
\begin{equation}
\f12\leq \f{a}{c}\leq \f32 
\end{equation}
is obeyed in the case of $n$ minimal punctures for all values of the parameters. 

From now on without loss of generality we take  $\epsilon_{\text{max}}>0$. We can construct all relevant and marginal operators solely from $\mu$ and $\phi$. Every $\mathcal{N}=2$ node is associated to two marginal operators, $\Tr \mu_i \phi_i$ and $\Tr\mu_{i+1}\phi_i$. If the gauge node connects two blue matter fields (i.e. $\sigma_i=-1$) it has two relevant operators, $\Tr\mu_{i}^2$ and  $\Tr\mu_{i+1}^2$, associated to it. If it connects two red matter fields (i.e. $\sigma_i=1$) there is only one relevant operator $\Tr\phi_i^2$. At an $\mathcal{N}=1$ node we have a single marginal operator, $\Tr\mu_{i}\mu_{i+1}$, and a single relevant one, $\Tr \mu_i^2$, where the adjacent blue matter field is labelled with $i$. Furthermore, we can construct gauge invariant operators out of the tri-fundamental fields $Q$ and $\widetilde{Q}$. These operators correspond to the wrapped M2-brane operator considered in Section~\ref{sec:punctures} and are given by
\begin{equation}
\mathcal{O}_{M2} = \prod_{i=1}^{2g-2}\prod_{j=1}^{n}Q_i\mu_j^+\,,\qquad \widetilde{\mathcal{O}}_{M2} = \prod_{i=1}^{2g-2}\prod_{j=1}^{n}\widetilde{Q}_i\mu_j^-\,.
\end{equation}
These operators have dimensions
\begin{equation}
	\Delta(\mathcal{O}_{M2})=\Delta(\widetilde{\mathcal{O}}_{M2}) =\f34(N-1)(p+q+\epsilon(p-q))+\f32(x+y+\epsilon(x-y))\,, 
\end{equation}
For large $N$ this dimension exactly matches the energy of the wrapped brane computed in \eqref{dimM2}. This provides a further consistency check of our construction.

Using the results above we can also compute the dimension of the $\mathcal{N}=1$ conformal manifold using the strategy of \cite{Leigh:1995ep,Green:2010da}. Our theory contains $3\mathbf{g}-3+n$ gauge couplings and $v_1 + 2 v_2$ additional marginal operators from the vector multiplets, giving a total of $3\mathbf{g}-3+n+v_1+2v_2$ marginal deformations. However, there are constraints on the anomalous dimensions coming from each of the $2\mathbf{g}-2$ $\mathcal{T}_N$ blocks and from all $\mathcal{N}=2$ vectors with the exception of one overall combination. Thus we find the dimension of the conformal manifold
\begin{equation}
\dim_\mathbf{C}\mathcal{M}_C = \mathbf{g} + n + v_1 + v_2 = 4\mathbf{g} - 3 + 2n
\end{equation}
which matches the counting in the dual gravity solutions \eqref{dimMc}.

Additionally we can deform our theory with relevant operators corresponding to mass terms for the adjoint chiral. In \cite{Tachikawa:2009tt} it was shown that when an $\mathcal{N}=2$ SCFT with a marginal coupling is deformed by adding mass terms for all the adjoint chirals in the vector multiplet it flows to a $\mathcal{N}=1$ SCFT and the central charges of the IR theory are related to those of the UV theory by a universal linear transformation
\begin{equation}\label{universalac}
a_{\rm IR} = \f{9}{32}(4 a_{\rm UV}-c_{\rm UV})\,,\qquad c_{\rm IR} = \f{1}{32}(-12 a_{\rm UV}+39 c_{\rm UV})\,.
\end{equation} 
In the large $N$ limit this implies that the ratio of the central charges is given by
\begin{equation}
\f{a_{\rm IR}}{a_{\rm UV}} = \f{c_{\rm IR}}{c_{\rm UV}} = \f{27}{32}\,.
\end{equation}
We observe that indeed the relations \eqref{universalac} are satisfied by our quiver gauge theories provided that the UV theory is the one with $\mathfrak{z}=1$ and the IR theory is the one with $\mathfrak{z}=0$ for fixed $N$ and $n$. These are exactly the two cases where $a$-maximization was not needed as a result of which the central charges are rational. At the universal IR point we have $\epsilon=0$ and the superconformal $R$ symmetry is simply $R_0$. At this point there are no relevant operators left solidifying its status as the inevitable universal IR fixed point. At this point the flavor symmetry is enhanced to $\SU(2)_\mathcal{F}$. Using the same arguments as in \cite{Benini:2009mz} we can show that the dimension of the conformal manifold in this case becomes 
\begin{equation}
	\dim_\mathbf{C} \mathcal{M}_C = 6g-6+3n\,.
\end{equation}

\begin{figure}[H]
	\centering	
	\begin{minipage}{0.3\textwidth}
		a) 
		\centering
		\includegraphics[scale=0.5]{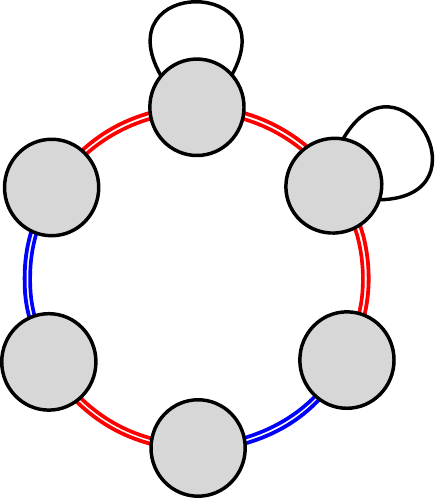}
	\end{minipage}
	\begin{minipage}{0.3\textwidth}
		b)
		\centering
		\includegraphics[scale=0.5]{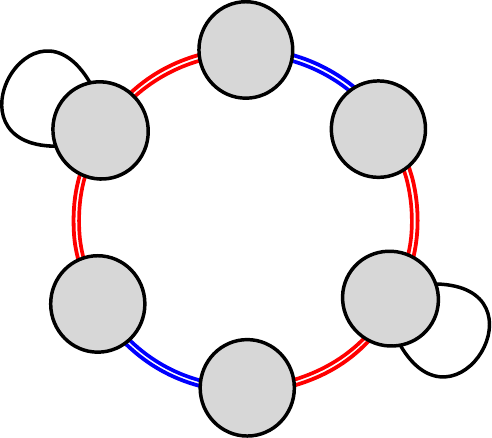}
	\end{minipage}
	\begin{minipage}{0.3\textwidth}
		c)
		\centering
		\includegraphics[scale=0.5]{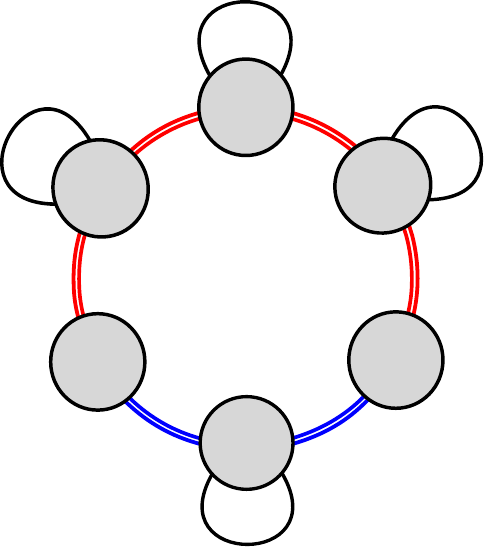}
	\end{minipage}
	\caption{All quivers with $\mathbf{g}=1$, $n=6$, $\mathfrak{z}=0$ and $\eta=\f{1}{3}$. The three quivers are related by Seiberg dualities.}
	\label{fig:k=6}
\end{figure}

\begin{figure}[H]
	\begin{minipage}{0.5\textwidth}
		\centering
		\includegraphics[scale=0.35]{quiver3}
	\end{minipage}
	\begin{minipage}{0.49\textwidth}
		\centering
		\includegraphics[scale=0.35]{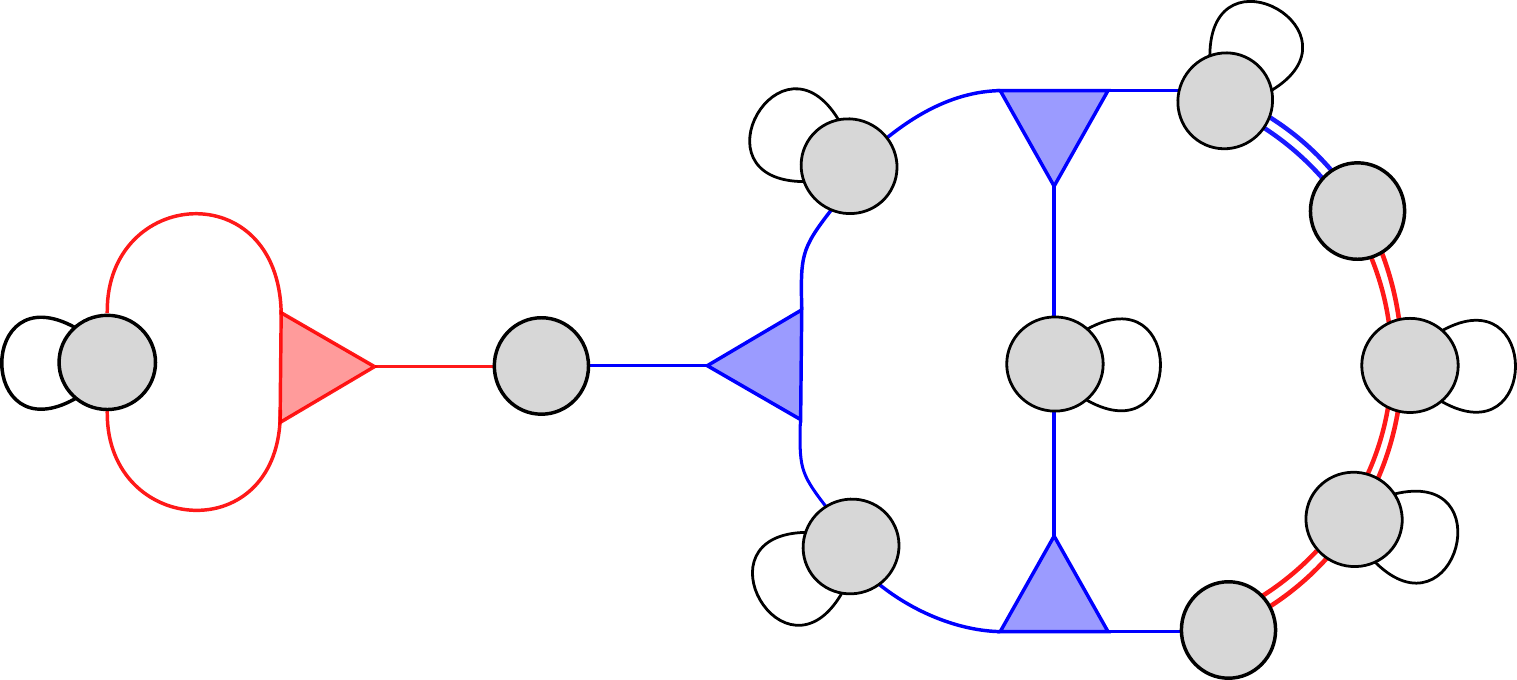}
	\end{minipage}
	\caption{An example of two different quivers which describe the same IR SCFT. The parameters specifying the quiver are $\mathbf{g}=3$, $z=\frac{1}{2}$, $\eta = -\frac{1}{2}$. The rank of the theory, $N$, is arbitrary.} 
	\label{fig:Seibergd}
\end{figure}

To finish this section, we briefly consider the various Seiberg duality transformations \cite{Seiberg:1994pq,Berenstein:2002fi} of our quiver gauge theories. Our SCFTs are labelled by four parameters, $\{ N, n, \mathfrak{z}, \eta \}$. For one possible set of parameters there may be multiple UV quivers realizing that particular set and we conjecture that they are all Seiberg dual. As a first example we consider the torus with $\mathfrak{z}=0$. For $n=6$ there are three quivers with $\eta=0$, three quivers with $\eta=\f{1}{3}$, one quiver with $\eta=\f{2}{3}$ and one quiver with $\eta=1$. All quivers with $n=6$ and $\eta=\f{1}{3}$ are depicted in Figure~\ref{fig:k=6}.
Performing a Seiberg duality on the rightmost $\mathcal{N}=1$ node of Figure~\ref{fig:k=6}a results in Figure~\ref{fig:k=6}b. On the other hand, performing a Seiberg duality on the lower $\mathcal{N}=1$ node of Figure \ref{fig:k=6}a results in Figure \ref{fig:k=6}c corroborating the conjecture that all quivers with the same parameters are Seiberg dual to each other. For higher genus Riemann surfaces there are even more intriguing Seiberg-like dualities where one can move vector multiplets across $\mathcal{T}_N$ blocks. An example is of this is presented in Figure~\ref{fig:Seibergd}
As noted in \cite{Bah:2012dg} for $\mathbf{g}>1$ quivers with no punctures, a naive counting of the relevant operators suggests a difference in the spectrum of chiral operators. However, when considering the $\mathcal{N}=1$ superconformal index \cite{Beem:2012yn,Gadde:2011uv,Rastelli:2014jja} of the two theories it has been shown that indeed the number of relevant operators is equal in both cases. The apparent difference in the spectrum of the chiral ring is due to non-trivial chiral ring relation that have to be properly taken into account. We thus conjecture that, in a similar vein, all quiver theories with the same discrete parameters we discussed here are dual to each other. In this way we uncover a multitude of interesting new Seiberg-like dualities similar to the ones studied in\cite{Gadde:2013fma}.

\section{Discussion} 
\label{sec:discussion}

The goal of this work was to show that one can use gauged supergravity in 4, 5, 6, and 7 dimensions to study the IR dynamics of M2-, D3-, D4-D8- and M5-branes wrapped on a singular complex curve $\mathcal{C}$. This is achieved by reducing the BPS equations of the supergravity theory to the Liouville equation on $\mathcal{C}$. Singular solutions to this equation are then interpreted in supergravity as holographic dual to the topologically twisted SCFTs living on the worldvolume of the branes. To provide evidence for our construction we described the details of this picture for the case of four-dimensional $\mathcal{N}=1$ SCFTs of class $\mathcal{S}$ arising from M5-branes wrapped on $\mathcal{C}$. There are several natural and important directions to pursue in order to ellucidate exploring these ideas and we discuss a few of them below.

Given the successful implementation of our approach to the physics of M5-branes wrapped on $\mathcal{C}$ it is natural to study in detail the physics of the other wrapped branes. The SCFTs dual to the known supergravity solutions corresponding to a smooth $\mathcal{C}$ are under much less control when compared to the class $\mathcal{S}$ theories we studied here. Indeed, this provides an instance where the supergravity analysis may inform the construction of the dual field theories. A particularly rich class of examples is offered by D3-branes wrapped on $\mathcal{C}$. When the Riemann surface is smooth the corresponding two-dimensional $(0,2)$ SCFTs were studied in \cite{Benini:2012cz,Benini:2013cda} and it will be most interesting to extend this to supergravity solutions with punctures on $\mathcal{C}$. Work along these lines is currently in progress \cite{BBGP}. Another important question is to study higher-curvature corrections to our supergravity solutions which should capture $1/N$ effects in the dual SCFTs. A possible starting point to attack this is provided by the construction in \cite{Baggio:2014hua}. Finally, it will be very interesting to generalize our construction to twisted compactifications of brane setups with smaller amount of supersymmetry, like the ones studied in \cite{Benini:2015bwz,Azzurli:2017kxo,Bobev:2017uzs}, or to other gauged supergravity theories arising as consistent truncations of string or M-theory, see \cite{Guarino:2015vca,Bah:2017wxp}.

Our construction provides a roundabout way to construct $\tfrac{1}{4}$-BPS AdS$_5$ solutions of eleven-dimensional supergravity with  explicit brane sources. These solutions should fall within the general classification of \cite{Gauntlett:2004zh}  of supersymmetric AdS$_5$ solutions in eleven dimensions. In fact, such backgrounds with explicit brane sources corresponding to punctured Riemann surface were studied in \cite{Bah:2013qya,Bah:2015fwa} and it is desirable to make the connection between that approach and our construction more explicit. A common feature between our setup and the one in \cite{Gaiotto:2009gz,Bah:2013qya,Bah:2015fwa} is that the punctures/singularities on the Riemann surface are introduced ``by hand'' as explicit singular sources in the BPS equations of the supergravity theory. It is desirable to make this more rigorous by adding explicit sources to the supergravity Lagrangian due to the presence of  branes in gauged supergravity. These arise from a dimensional reduction of the D- or M-branes in ten and eleven dimensions. The brane sources would allow us to derive more directly the BPS equations we used.

The gravity solutions discussed in Section~\ref{sec:M5branes} and Section~\ref{sec:punctures} allow for the line bundles over the Riemann surface to have negative degrees. In field theory discussion of Section~\ref{sec:quivers}, however, we restricted to positive degrees only. The quiver gauge theories dual to these negative degree line bundle setups can be constructed by introducing additional field theory building blocks $\mathcal{T}_N^{(m)}$ discussed in \cite{Agarwal:2015vla}. These building blocks can be obtained by adding adjoint chiral multiplets charged under the global symmetry of $\mathcal{T}_N$ and giving them a nilpotent vacuum expectation value, which in turn spontaneously breaks the global symmetry. In this way one can obtain more general quivers \cite{Fazzi:2016eec,Nardoni:2016ffl,Maruyoshi:2016aim} allowing for negative degrees of the line bundles. Furthermore, using these building blocks we can construct gauge theories on the torus which include background flux. It will be interesting to study these more general setups in detail. It should also be stressed that in this work we focused on quivers built out of $\SU(N)$ vector multiplets and $\SU(N)$ $\mathcal{T}_N$ theories only, i.e., theories describing the infrared dynamics of the $A_{N-1}$ type $\mathcal{N}=(2, 0)$ theory on $\mathcal{C}$. It is clear that these constructions could be generalized to quivers built out of $\SO(2N)$ and $\Sp(2N - 2)$ vector multiplets and the $\SO(2N)$ $\mathcal{T}_N$ theories to describe the infrared limit of the $D_N$ theory compactified on $\mathcal{C}$, see for instance \cite{Tachikawa:2009rb}. 

In Section~\ref{subsec:g1Laba} we noted that the $L^{aba}$ quiver gauge theories arising on the worldvolume of D3-branes probing a conical CY singularity can be constructed also from M5-branes wrapped on a torus with punctures. It is important to understand whether other quiver gauge theories in the $L^{abc}$ or $Y^{pq}$ can also be realized in terms of M5-branes in the spirit of class $\mathcal{S}$. The fact that the AdS$_5$ solutions associated to the $Y^{pq}$ metrics were first constructed in \cite{Gauntlett:2004zh} as (singular) solutions of eleven-dimensional supergravity suggests that such a connection may be possible.

In this work we focused on studying wrapped branes on a Riemann surface with punctures. A natural generalization is to study similar supergravity setups for branes wrapped on higher-dimensional calibrated cycles. When the cycles are smooth and compact it is well-known how to construct the corresponding supergravity solutions, see \cite{Gauntlett:2003di,Bobev:2017uzs} for a review. These constructions again typically involve studying the BPS equations of a lower-dimensional gauged supergravity. It will be interesting to apply our approach and introduce explicit singular sources to these BPS equations to study the supergravity solutions describing branes wrapped on manifolds with various defects. A particularly accessible example is offered by the case of M5-branes wrapping a four-cycle given by a product of two Riemann surfaces. The supergravity solutions corresponding to smooth surfaces were analyzed in detail in \cite{Benini:2013cda}, see also \cite{Bah:2015nva}, and it should be straightforward to apply our construction to this setup to construct solutions with punctures.

Given the appearance of the Liouville equation in our construction and the $\SU(\infty)$ equation in the supergravity analysis of \cite{Lin:2004nb,Gaiotto:2009gz} it is natural to wonder whether these equations are some remnant of the underlying AGT correspondence for M5-branes wrapped on $\mathcal{C}$.

\bigskip
\bigskip
\leftline{\bf Acknowledgements}
\smallskip
\noindent We are grateful to Ibrahima Bah, Anthony Charles, Emily Nardoni, Klaas Parmentier, and Shlomo Razamat for interesting discussions. The work of NB and PB is supported in part by the Odysseus grant G0F9516N from the FWO. FFG is a Postdoctoral Fellow of the Research Foundation - Flanders (FWO). This work is also supported in part by the KU Leuven C1 grant ZKD1118 C16/16/005.

\appendix

\section{Derivation of the BPS equations}
\label{app:BPS}

The gauged supergravity BPS equations for the Ansatz in Section \ref{sec:puncbranes} can be reduced to the Liouville equation on the Riemann surface accompanied by a set of algebraic equations. This is shown in detail in \cite{BGP} for gauged supergravity theories in four, five, six and seven dimensions. Here we provide a derivation of this for the case of main interest in this paper, namely seven-dimensional maximal gauged supergravity.

As in \cite{Anderson:2011cz,Bah:2011vv,Bah:2012dg} we work with the $\U(1)\times \U(1)$ invariant truncation of the seven-dimensional maximal $\SO(5)$ gauged supergravity of \cite{Pernici:1984xx}. The supersymmetry variations for the fermionic fields were derived in \cite{Pernici:1984xx}. For the truncation of interest here they read \cite{Liu:1999ai} 
\begin{equation}\label{AAEq3}
\begin{aligned}
\delta\psi_{\mu} =& \left[ \nabla_{\mu} +2g(A_{\mu}^{(1)}\Gamma^{12} + A_{\mu}^{(2)}\Gamma^{34}) + \frac{g}{2} e^{-4(\lambda_1+\lambda_2)}\gamma_{\mu} + \frac{\gamma_{\mu}}{2} \gamma^{\nu} \partial_{\nu} (\lambda_1+\lambda_2) \right] \varepsilon\\
& \qquad\qquad+ \frac{\gamma^{\nu}}{2} \left( e^{-2\lambda_1}F_{\mu\nu}^{(1)}\Gamma^{12} + e^{-2\lambda_2}F_{\mu\nu}^{(2)}\Gamma^{34} \right) \varepsilon~, \\
\delta\chi^{(1)} =& \left[ \frac{g}{2} (e^{2\lambda_1}- e^{-4 (\lambda_1+\lambda_2)}) - \frac{\gamma^{\mu}}{4}  \partial_{\mu} (3\lambda_1+2\lambda_2) - \frac{\gamma^{\mu\nu}}{8}e^{-2\lambda_1}F_{\mu\nu}^{(1)}\Gamma^{12} \right] \varepsilon ~, \\
\delta\chi^{(2)} =& \left[ \frac{g}{2} (e^{2\lambda_2}- e^{-4 (\lambda_1+\lambda_2)}) - \frac{\gamma^{\mu}}{4}  \partial_{\mu} (2\lambda_1+3\lambda_2) - \frac{\gamma^{\mu\nu}}{8}e^{-2\lambda_2}F_{\mu\nu}^{(2)}\Gamma^{34} \right] \varepsilon ~, 
\end{aligned}
\end{equation}
where $g$ is the gauge coupling of the supergravity theory which is related to the radius of the maximally symmetric AdS$_7$ solution. The $\Gamma^i$ are $\SO(5)$ gamma matrices, the $\gamma_{\mu}$ seven-dimensional space-time gamma matrices, and we have defined $\gamma_{\mu_1\ldots\mu_p} = \gamma_{[\mu_1}\ldots\gamma_{\mu_p]}$. From now on we suppress all spinor indices and use hats to indicate tangent space indices. The partial topological twist of the boundary field theory suggests the following projectors,
\begin{equation}
\gamma_{\hat{x}_1\hat{x}_2}\varepsilon={\rm i}\varepsilon~,\qquad \Gamma^{12}\varepsilon={\rm i}\varepsilon~,\qquad \Gamma^{34}\varepsilon=i\varepsilon~,\qquad \gamma_{\hat r}\varepsilon=\varepsilon\,.
\label{spinorcharges}
\end{equation}
Therefore, in general, our solutions preserve one quarter of the supersymmetry. Four-dimensional Poincar\'e invariance implies that the spinors are constant in the $\mathbf{R}^{1,3}$ directions,
\begin{equation}
\partial_{t}\varepsilon=\partial_{z_i}\varepsilon=0~.\label{AAEq8}
\end{equation}
Note however that we do not assume that the spinors are independent of the coordinates $(x_1,x_2)$ parametrizing the Riemann surface. The conditions for the supersymmetry variations \eqref{AAEq3} to vanish are of two types. Vanishing of the variation of the dilatinos $\chi^{(i)}$ and the $(t,z_1,z_2,z_3)$ components of the gravitino $\psi_{\mu}$ leads to differential equations for the background fields, while the integrability of this system imposes additional constraints. One can show that for the truncation of interest and with the Ansatz in \eqref{Ansatz}-\eqref{IRfields} these reduce to: 
\begin{equation}\label{BPSall}
\begin{aligned}
&  \partial_{r}(3\lambda_1+2\lambda_2) - 2g e^{f+2\lambda_1} +2g e^{f-4\lambda_1-4\lambda_2} -  e^{h - \hat{\varphi} - 2\lambda_1} F^{(1)}_{x_1x_2}=0~,\\
&  \partial_{r}(2\lambda_1+3\lambda_2) - 2g e^{f+2\lambda_2} +2g e^{f-4\lambda_1-4\lambda_2} -  e^{f - \hat{\varphi} - 2\lambda_2} F^{(2)}_{x_1x_2}=0~,\\
& (\partial_{x_1}+\rmi\partial_{x_2})(3\lambda_1+2\lambda_2) - e^{-f - 2\lambda_1 }(F^{(1)}_{x_2r}-\rmi F^{(1)}_{x_1r})=0~,\\
& (\partial_{x_1}+\rmi\partial_{x_2})(2\lambda_1+3\lambda_2) - e^{-f - 2\lambda_2 }(F^{(2)}_{x_2r}-\rmi F^{(2)}_{x_1r})=0~,\\
& \partial_{r}\left(f +\lambda_1+\lambda_2\right) + g e^{f-4\lambda_1-4\lambda_2} = 0~, \\
& \partial_{x_1}\left(f +\lambda_1+\lambda_2\right) = \partial_{x_2}\left(f +\lambda_1+\lambda_2\right) = 0~,\\
& \partial_{r}(\hat{\varphi}/2 - 4\lambda_1 - 4\lambda_2) +  2g e^{f +2\lambda_1} +  2g e^{f +2\lambda_2} - 3g e^{f - 4\lambda_1 - 4\lambda_2} =0 ~,\\
& \partial_{r}\partial_{x_2}(\hat{\varphi}/2-4\lambda_1-4\lambda_2) + 4g F^{(1)}_{rx_1}+ 4g F^{(2)}_{rx_1} = 0~,\\
& \partial_{r}\partial_{x_1}(\hat{\varphi}/2-4\lambda_1-4\lambda_2) + 4g F^{(1)}_{x_2r}+ 4g F^{(2)}_{x_2r} = 0~, \\
& (\partial_{x_1}^2 + \partial_{x_2}^2)(\hat{\varphi}/2-4\lambda_1-4\lambda_2) - 4g F^{(1)}_{x_1x_2} - 4g F^{(2)}_{x_1x_2}=0~. 
\end{aligned}
\end{equation}
Imposing that the variations of the $(r,x_1,x_2)$ components of the gravitino vanish implies the equation 
\begin{equation}
	\partial_r \epsilon - \tfrac{1}{2} (\partial_r f)\epsilon = 0\,,
\end{equation}
which is solved by $\epsilon = \e^{f/2}\epsilon_0$ with $\epsilon_0$ a constant spinor obeying the projectors in \eqref{spinorcharges}. In addition these gravitino equations imply that the two $\U(1)$ gauge fields are given by
\begin{equation}\label{vectwist}
	A^{(1)} = \frac{a_1}{8g(a_1+a_2)} \dd\varphi\,, \qquad A^{{(2)}} = \frac{a_1}{8g(a_1+a_2)} \dd\varphi\,.
\end{equation}
Here we show that in the IR, where the metric takes the form AdS$_5\times\mathcal{C}$ these equation reduce to a single second order equation for the conformal factor $\varphi$ together with algebraic equations for the other fields. Plugging the IR fields \eqref{IRfields} in \eqref{BPSall} results in the following equations
\begin{equation}
\begin{aligned}
&  2g e^{2\lambda_1} -2g e^{-4\lambda_1-4\lambda_2} +  e^{- \varphi -\varphi _0- 2\lambda_1} F^{(1)}_{x_1x_2}=0~, \\
& 2g e^{2\lambda_2} -2g e^{-4\lambda_1-4\lambda_2} + e^{ -\varphi -\varphi _0 - 2\lambda_2} F^{(2)}_{x_1x_2}=0~, \\
& 1 - 2 e^{f_0-4\lambda_1-4\lambda_2} = 0~, \\
&e^{2\lambda_1} + e^{2\lambda_2} - \f32 e^{- 4\lambda_1 - 4\lambda_2} =0 ~, \\
& (\partial_{x_1}^2 + \partial_{x_2}^2)\varphi - 8g (F^{(1)}_{x_1x_2}+ F^{(2)}_{x_1x_2})=0~.
\end{aligned}
\end{equation}
The third and fourth of these equations, together with a linear combination of the first two lead to algebraic equations for the scalars, and the metric constants. Finally, after defining, 
\begin{equation}
\e^{\varphi_0} = \f{\e^{4\lambda_1+4\lambda_2}}{16g^2}(\e^{8\lambda_1+4\lambda_2}+\e^{4\lambda_1+8\lambda_2}-\e^{-2\lambda_1}-\e^{2\lambda_2})^{-1}
\end{equation}
and combining the second linearly independent combination, together with the last equation we find the Liouville equation 
\begin{equation}\label{eq:Liouvilleeqn}
\square \varphi + \kappa \e^\varphi =(\partial_{x_1}^2 + \partial_{x_2}^2)\varphi +\kappa \e^\varphi = 0\,,
\end{equation}
which has to be satisfied by the conformal factor $\varphi$.

\section{The Liouville equation}\label{app:Liouville}

Here we give a short overview of the Liouville equation and some of its properties that are relevant in the context of this work. The Liouville equation is the non-linear partial differential equation \eqref{eq:Liouvilleeqn} satisfied by the conformal factor $\varphi$ of a metric $\dd s^2 = \e^\varphi(\dd x_1^2 + \dd x_2^2)$ on a surface of constant Gaussian curvature $\kappa$.\footnote{Note that it is sometimes convenient to work in complex coordinates $z=\tfrac{1}{2}(x_1+\rmi x_2)$ where $\square = \partial_{x_1}^2 + \partial_{x_2}^2 = 4\partial_z\partial_{\bar{z}}$.} This equation can be used to prove the uniformization theorem which states that every simply connected Riemann surface is conformally equivalent to one of three Riemann surfaces:
\begin{itemize}
	\item the hyperbolic plane for $\kappa<0$,
	\item the complex plane for $\kappa = 0$,
	\item the Riemann sphere for $\kappa >0$.
\end{itemize}
In particular this implies that every Riemann surface admits a Riemannian metric of constant curvature. For compact Riemann surfaces, the hyperbolic Riemann surfaces with genus $\mathbf{g}>1$ have the hyperbolic plane as universal cover and have a non-abelian fundamental group. The torus, $\mathbf{g}=1$ has the complex plane as universal cover and the fundamental group is $\mathbf{Z}^2$. Finally the Riemann sphere with genus $\mathbf{g}=0$ has a trivial fundamental group. 

Apart from regular solutions to the Lioville equation we also consider Riemann surfaces with prescribed singularities. This results in the addition of localized sources on the right hand side of the Liouville equation
\begin{equation}\label{appendixLiuwsources}
\square \varphi + \kappa \e^\varphi = 4\pi\sum_{i}(1-\xi_i)\delta^{(2)}(p_i)\,,
\end{equation}
where $2\pi\xi_i$ parametrizes the opening angle of the conical defects located at the point $p_i$ and we restrict the parameters to lie within the interval $0<\xi_i<1$. We can specify a Riemann surface of interest by specifying a set of parameters $\{ \mathbf{g}, J, \vec{\xi}, \vec{p} , \kappa \}$, where $\mathbf{g}$ is the genus, $J$ a complex structure, $\vec{\xi}$ and $\vec{p}$ two vectors with entries the opening angles and positions of the conical singularities and $\kappa$ the Gaussian curvature. Given a metric $g$ on the Riemann surface, we can compute the volume using the Gauss-Bonnet theorem.
\begin{equation}\label{volriem}
	\vol_{\mathbf{g},\xi}  = \int_\Sigma \kappa_g \dd \omega_g =  \f{2\pi}{\kappa} \chi(\Sigma,\vec{\xi}) = \f{2\pi}{\kappa}\left( \chi(\Sigma) - \sum_i(1-\xi_i) \right)\,,
\end{equation}
where $\chi(\Sigma) = 2g-2$ is the topological Euler characteristic and we call $\chi(\Sigma,\vec{\xi})$ the conic Euler characteristic. In algebraic geometry language this represents the twisted anticanonical divisor, $-K_\Sigma -\sum_i(1-\xi_i)p_i$ on $\Sigma$, where $K_\Sigma = T^{1,0*}\Sigma$ denotes the class of the canonical divisor of the smooth Riemann surface $\Sigma$, see for example \cite{McOwen:1988,Troyanov:1991,Luo:1992}. When a constant curvature metric exists the sign of the curvature $\kappa_g$ agrees with the sign of $\chi(\Sigma,\vec{\xi})$. To extend the uniformization theorem to include Riemann surfaces with conical singularities, one looks for a metric compatible with the complex structure $J$, with conical singularities at points $p_i$ and constant curvature away from the singularities. 
Indeed, using \eqref{appendixLiuwsources} it follows that the Ricci scalar, given by $R = -\e^{-\varphi}\square\varphi$, is constant away from the singularities. Near a conical singularity, $\varphi \propto \log|z|$ implying that at these points $-\square\varphi$ is given by a multiple of delta functions at $z=0$.\footnote{This can be seen by excising a small circle around the origin and invoking the Stokes theorem}. In \cite{McOwen:1988,Troyanov:1991,Luo:1992} it was shown that such a constant curvature metric always exists when $\chi(\Sigma,\vec{\xi})\leq 0$ or when $\chi(\Sigma,\vec{\xi})>0$ and $1-\xi_i > \sum_{j\neq i}(1-\xi_j)$ for all $i$. This metric is furthermore unique except when $\chi(\Sigma,\vec{\xi})=0$ when it is unique up to an overall constant, or when $\Sigma = S^2$ with less than three punctures when it is unique up to a M\"obius transformation which fixes the position of the singularities. It is known that to each such set of parameters there exists a unique metric on $\Sigma$ with constant Gaussian curvature $\kappa$ and prescribed singularities at the points $p_i$.

Away from the singularities, the most general solution to the Liouville equation is given by
\begin{equation}\label{gensol}
	\varphi = \log\left(4\f{|\partial_z u(z)|^2}{(1+\kappa |u(z)|^2)^2}\right)\,,
\end{equation}
where $u(z)$ is a meromorphic function with non-vanishing holomorphic derivative and at most simple poles. Close to the punctures, the Liouville equation implies the following asymptotic behaviour:
\begin{equation}
\varphi = -2(1-\xi_i)\log|z-z_i|\,, \qquad \text{as } z\rightarrow z_i\,.
\end{equation}
The problem of finding a general solution with conical singularities is closely related to the Riemann-Hilbert problem of finding functions with prescribed monodromies in the complex plane. In order to solve this problem we introduce the Fuchsian equation, given $n$ singularities with opening angles $\xi_i$ this equation is given by
\begin{equation}\label{fuchs}
	\f{\dd^2 w}{\dd z^2} + \sum_{i=1}^{n}\left[\f{(1-\xi_i)(1+\xi_i)}{4(z-z_i)^2} + \f{c_i}{2(z-z_i)}\right] w = 0\,,
\end{equation}
where the $c_i$ are known as the accessory parameters which have to satisfy the equations
\begin{equation}
\left\{\begin{array}{l}
\sum_{i}c_i=0\\
\sum_i(2c_iz_i + (1-\xi_i)(1+\xi_i))=0\\
\sum_i(c_i z_i^2+z_i(1-\xi_i)(1+\xi_i))=0\,.
\end{array}\right.
\end{equation}
The double poles of \eqref{fuchs} fix the behaviour of the solutions near the singular points to
\begin{equation}
	w(z) \sim A (z-z_i)^{(1+\xi_i)/2} + B (z-z_i)^{(1-\xi_i)/2}\,,
\end{equation}
from which one can easily read of the monodromies. Given a pair of linearly independent solutions $w_1$ and $w_2$ one can see that by plugging the function $u=w_1/w_2$ in \eqref{gensol} we find a solution $\varphi$ to the Liouville equation with the prescibed singularities. In general the monodromies belong to $\SL(2,\mathbf{C})$, and $\varphi$ will not be a single valued function. In order to find a single valued function we need to furthermore require all monodromies to lie in $\SU(2)$, or $\SU(1,1)$ for $\kappa<1$. These conditions uniquely determine the accessory parameters $c_i$ and consequently the function $w_{1,2}$. Since now $u$ transforms as 
\begin{equation}
	u\rightarrow \frac{a u + b}{-\bar{b}u + \bar{a}}\,,\qquad \text{where } |a|^2+ |b|^2 = 1\,,
\end{equation}
all monodromies leave $\varphi$ invariant.

\section{Uplift formulae}\label{app:uplift}

In this appendix we collect all the relevant uplift formulae used in this paper. The uplift formulae for the maximal four, five and seven-dimensional supergravities to string and/or M-theory were given in \cite{Cvetic:1999xp}. The uplift formulae from the maximal six-dimensional supergravity to ten-dimensional type IIA supergravity was given in \cite{Cvetic:1999un}.

\subsection{$S^7$ reduction of eleven-dimensional supergravity}
\label{app:upliftM2}

The eleven-dimensional metric is given by
\begin{equation}
\dd s_{11}^2 = \Delta^{2/3}\dd s_4^2 + \f{\Delta^{-1/3}}{g^2}\sum_{i=1}^{4}X_i^{-1}\left( \dd\mu_i^2 + \mu_i^2\left(\dd\phi_i+g A^{(i)}\right)^2 \right)\,.
\end{equation}
Here the four functions $\mu_i$ satisfy the constraint $\sum_i\mu_i^2=1$ and the $X_i$ are defined as
\begin{equation}
\begin{aligned}
&X_1 = \e^{\f14(3\lambda_1-\lambda_2-\lambda_3)}\,,\qquad\qquad X_2 = \e^{\f14(3\lambda_2-\lambda_1-\lambda_3)}\,,\\
&X_3 = \e^{\f14(3\lambda_3-\lambda_1-\lambda_2)}\,,\qquad\qquad X_4 = \e^{\f14(-\lambda_1-\lambda_2-\lambda_3)}\,.
\end{aligned}
\end{equation}
A convenient parametrization for the $\mu_i$ is given by
\begin{equation}
\begin{aligned}
&\mu_1 = \sin\alpha\,,\qquad\qquad\hspace{18pt}\qquad \mu_2 = \cos\alpha\sin\beta\,,\\
&\mu_3 = \cos\alpha\cos\beta\sin\gamma\,,\quad\qquad \mu_4 = \cos\alpha\cos\beta\cos\gamma\,. 
\end{aligned}
\end{equation}
The function $\Delta$ appearing in the metric is given by
\begin{equation}
\Delta = \sum_i X_i \mu_i^2\,.
\end{equation}
There is also a four-form field strength, given by
\begin{multline}
F_{(4)} = 2g\sum_i\left(X_i^2 \mu_i^2-X_i\Delta\right)\epsilon_{(4)}\\ -\f{1}{2g^2}\sum_i X_i^{-2}\dd(\mu^2_i)\wedge \left( \dd\phi_i + g A^{(i)} \right)\wedge \star_4 F^{(i)}\,. 	
\end{multline}
Here $\epsilon_{(4)}$ is the volume form on the four-dimensional part of the metric $\dd s_4^2$.

\subsection{$S^5$ reduction of type IIB supergravity}
\label{app:upliftD3}

The uplift to type IIB supergravity is given by
\begin{equation}
\dd s_{10}^2 = \Delta^{1/2}\dd s^2_{5} + \f{ \Delta^{-1/2}}{g^2}\sum_{i=1}^{3}X_i^{-1}\left( \dd\mu_i^2 + \mu_i^2\left(\dd\phi_i+g A^{(i)}\right)^2 \right)
\end{equation}
where $\mu_i$ are subject to the constraint $\sum\mu_i^2=1$. The $X_i$ are given by
\begin{equation}
X_1 = \e^{-\lambda_1-\lambda_2} \,, \qquad X_2 =\e^{-\lambda_1+\lambda_2} \,, \qquad X_3 = \e^{2\lambda_1}\,.
\end{equation}
A convenient parametrisation for the $\mu_i$ is given by
\begin{equation}
\mu_1 = \cos\alpha\sin\beta\,,\qquad \mu_2 = \cos\alpha\cos\beta\,,\qquad \mu_3 = \sin\alpha\,.
\end{equation}
The function $\Delta$ is given by
\begin{equation}
\Delta = \sum_{i=1}^{3}X_i \mu_i^2\,.
\end{equation}
The self-dual five-form field strength is given by $F_5 = G_5 + \star_{10}G_5$ with
\begin{equation}
G_5 = 2g \sum_i\left( X_i^2\mu_i^2 - \Delta X_i \right)\epsilon_5 - \f{1}{2g^2}\sum_i X_i^{-2} \dd \left(\mu_i^2\right)\wedge\left(\dd\phi_i+g A^{(i)}\right)\wedge \star_5 F^{(i)}
\end{equation}
where $\epsilon_5$ is the volume form on $\dd s_5^2$ and $\star_5$ is the five-dimensional Hodge dual with respect to the same metric.

\subsection{$S^4/\mathbf{Z}_2$ reduction of massive type IIA supergravity}
\label{app:upliftD4D8}

In ten dimensions, the metric is given by
\begin{equation}
\dd s_{10}^2 = (\sin\alpha X^{\f32})^{\tfrac{1}{12}}\Delta^{\f38}\left[ \dd s_6^2 + \f{2 X^2}{g^2}  \dd\alpha^2 + \f{1}{2g^2X\Delta} \cos^2\alpha \left(\sigma_1^2+\sigma_2^2+(\sigma_3-g A)^2\right)\right]\,.
\end{equation}
The only non-vanishing field strength is given by
\begin{multline}
F_{(4)} = -\f{\sqrt{2}\sin^{\f13}\alpha\cos^3\alpha}{6g^3} \Delta^{-2}U \dd\alpha\wedge\epsilon_{(3)} + \f{\sin^{\f13}\alpha\cos\alpha}{\sqrt{2}g^2}F\wedge (\sigma_3-g A)\wedge \dd \alpha\\
+\f{\sin^{\f43}\alpha\cos^2\alpha}{2\sqrt{2}g^2}\Delta^{-1}X^{-3}F\wedge \dd\sigma_3
\end{multline}
Finally the dilaton is given by
\begin{equation}
\e^\phi = \sin^{-\f56}\alpha \Delta^{\f14}X^{-\f54}\,.
\end{equation}
In these expressions we have introduced the functions
\begin{equation}
\begin{aligned}
\Delta =& X \cos^2\alpha + X^{-3}\sin^2\alpha\,,\\
U =& X^{-6}\sin^2\alpha+\left(4X^{-2}-3X^2\right)\cos^2\alpha - 6X^{-2}\,.
\end{aligned}
\end{equation}
The $\sigma_i$ are left-invariant one-forms of $\SU(2)$ which satisfy $\dd\sigma_i = -\f12\epsilon_{ijk}\sigma_j\wedge\sigma_k$. The gauge coupling constant $g$ is related to the mass parameter $M$ of the massive type IIA theory by $M = \f{\sqrt{2}}{3}g$.

\subsection{$S^4$ reduction of eleven-dimensional supergravity}
\label{app:upliftM5}

The eleven-dimensional metric is given by
\begin{equation}
\dd s_{11}^2 = \Delta^{1/3}\dd s_7^2 + \f{1}{4g^2}\Delta^{-2/3}\left[ X_0^{-1}\dd\mu_0^2+\sum_{i=1}^{2}X_i^{-1}(\dd\mu_i^2+\mu_i^2(\dd\phi_i + 2m A^{(i)})^2)\right]\,.
\end{equation}
The four-form flux takes the form
\begin{multline}
\star_{11}F_4 = 4g \sum_{i=0}^{2}\left(X_i^2\mu_i^2-\Delta X_i\right)\epsilon_{(7)} + 2g\Delta X_0 \epsilon_{(7)} \\+ \f{1}{4g^2}\sum_{i=1}^{2}X_i^{-2}\dd(\mu_i^2)\wedge (\dd\phi_i+4g A^{(i)})\wedge \star_7 F^{(i)}\,.
\end{multline}
Here we have introduced the functions $X_{0,1,2}$:
\begin{equation}
\begin{aligned}
X_1 = \e^{2\lambda_1}\,,\qquad X_2 = \e^{2\lambda_2}\,,\qquad X_0 = (X_1X_2)^{-2}\,.
\end{aligned}
\end{equation}
and
\begin{equation}
\Delta = \sum_{i=0}^{2}X_i \mu_i^2\,,
\end{equation}
while the $\mu_i$ are constrained to lie on the hypersurface $\sum_{i=0}^{2}\mu_i  =1$. A convenient parametrization for the $\mu_i$ is given by
\begin{equation}
\mu_0 = \cos\alpha\cos\beta\,,\qquad \mu_1 = \sin\alpha\,,\qquad \mu_2 = \cos\alpha\sin\beta\,.
\end{equation}
%

\section{Four-dimensional SCFTs}
\label{app:SCFTtrivia}

Here we collect some of our conventions for four-dimensional SCFTs. A four-dimensional $\mathcal{N}=2$ SCFT has $\SU(2)_R \times \U(1)_r$ R-symmetry. We denote the generator of the diagonal Cartan of $\SU(2)_R$ by $I_3$ and the generator of $\U(1)_r$ by $R_{\mathcal{N}=2}$. The charge assignments for the components of a $\mathcal{N}=2$ vector and hypermultiplet are given as follows:
\begin{figure}[H]
	\centering
\begin{minipage}{0.40\textwidth}
	\centering
	\begin{tabular}{c|ccc}  
		$R_{\mathcal{N}=2}\setminus I_3$ & $\frac{1}{2}$ & $0$ & $-\frac{1}{2}$ \\
		\hline 
		$0$ &  & $A_\mu$ & \\
		$1$ & $\lambda$ &  & $\lambda'$ \\ 
		$2$ & & $\phi$ &     
	\end{tabular}
\end{minipage}
\begin{minipage}{0.40\textwidth}
	\centering
\begin{tabular}{c|ccc}  
$R_{\mathcal{N}=2} \setminus I_3$ & $\frac{1}{2}$ & $0$ & $-\frac{1}{2}$ \\ 
\hline 
$-1$ &  & $\psi$ & \\ $0$ & $Q$ &  & $\widetilde{Q}^\dagger$ \\ 
$1$ & & $\widetilde{\psi}^\dagger$ &     
\end{tabular}
\end{minipage}
\end{figure}
Since we study both $\mathcal{N}=1$ and $\mathcal{N}=2$ theories, it is useful to consider an $\mathcal{N}=1$ subalgebra of the $\mathcal{N}=2$ algebra. A choice of subalgebra corresponds to a choice of Cartan of $\SU(2)_R$. The unique $\mathcal{N}=1$ superconformal R-symmetry generator in an $\mathcal{N}=2$ SCFT is given by
\begin{equation}
R_{\mathcal{N}=1} = \f13 R_{\mathcal{N}=2}+\f43 I_3\,.
\end{equation}
The linear combination 
\begin{equation}
	J = R_{\mathcal{N}=2}-2I_3\,,
\end{equation}
commutes with the chosen $\mathcal{N}=1$ subalgebra and is thus a flavour symmetry from the $\mathcal{N}=1$ point of view.

Finally we note that the operators in a $\mathcal{T}_N$ theory have charges
\begin{table}[H]
	\centering
	\begin{tabular}{c|ccc}
 	& $R_{\mathcal{N}=2}$ & &  $I_3$ \\
 	\hline
	$u_k$ & $2k$ &  & $0$ \\
	$Q$ & $0$  &  & $\frac 1 2 (N-1)$ \\
	$\widetilde Q$ & $0$  &  & $\frac12 (N-1)$ \\
	$\mu$ & $0$ &  & $1$ \\
\end{tabular}
\end{table} 
%


\bibliography{refs}
\bibliographystyle{JHEP}

\end{document}